\documentclass[twocolumn]{HeadersAndConfig/webofc}
\usepackage[varg]{txfonts}   
\usepackage[hidelinks]{hyperref}
\usepackage{comment}
\usepackage{fancyhdr}
\usepackage{graphicx}
\usepackage{braket}
\usepackage{multirow}
\usepackage{url}
\usepackage{hvfloat}
\usepackage{commath}
\usepackage{amsmath}
\usepackage{caption}
\usepackage{booktabs}
\usepackage{colortbl}
\usepackage{comment}
\usepackage[export]{adjustbox}
\usepackage{wrapfig}
\usepackage{csquotes}
\usepackage{upgreek}

\newlength{\bibitemsep}
\newlength{\bibparskip}
\let\oldthebibliography\thebibliography
\renewcommand\thebibliography[1]{%
  \oldthebibliography{#1}%
  \setlength{\parskip}{\bibitemsep}%
  \setlength{\itemsep}{\bibparskip}%
}

\usepackage[symbol,bottom]{footmisc}

\newcommand{\xmax}{$X_{\text{max}}$}
\newcommand{\xmaxbf}{$\mathbf{X_{\text{max}}}$}
\newcommand{\xmaxnorm}{$X_{\text{max}}^\prime$}

\newcommand{\xmaxmu}{$\langle X_{\text{max}} \rangle$}

\newcommand{\xmaxmunorm}{$\langle X_{\text{max}}^\prime \rangle$}

\newcommand{\xmaxsigma}{$\sigma( X_{\text{max}} )$}

\newcommand{\xmaxsigmanorm}{$\sigma ( X_{\text{max}}^\prime )$}

\newcommand{\lge}{$\log_{10}(E/\text{eV})$}

\newcommand{\gcm}{g/cm$^2$}

\newcommand{\Dxmaxmu}{$\Delta \langle X_{\text{max}} \rangle$}
\newcommand{\Dxmaxmunorm}{$\Delta \langle X_{\text{max}}^{\prime} \rangle$}
\newcommand{\Dxmaxsigma}{$\Delta\,\sigma( X_{\text{max}})$}
\newcommand{\Dxmaxsigmanorm}{$\Delta \sigma( X_{\text{max}}^{\prime} )$}

\newcommand{\myparagraph}[1]{\paragraph{#1}}

\newenvironment{myquote}[1]%
  {\list{}{\leftmargin=#1\rightmargin=#1}\item[]}%
  {\endlist}
\begin{document}
\title{Update on the indication of a mass-dependent anisotropy above 10\textsuperscript{18.7}\,eV in the hybrid data of the Pierre Auger Observatory }

\author{
    \firstname{Eric} \lastname{Mayotte}\inst{1,2}\fnsep\thanks{\email{emayotte@mines.edu}}
    \and
    \firstname{Thomas} \lastname{Fitoussi}\inst{3}\
    for the \lastname{Pierre Auger Collaboration}\inst{4}\fnsep\thanks{\email{spokespersons@auger.org}}
}

\institute{
    Colorado School of Mines, Department of Physics, Golden, CO, USA
    \and
    Bergische Universit\"at Wuppertal, Department of Physics, Wuppertal, Germany
    \and
    Karlsruhe Institute of Technology (KIT), Institute for Astroparticle Physics, Karlsruhe, Germany
    \and
    Observatorio Pierre Auger, Av.\ San Mart\'in Norte 304, 5613, Malarg\"ue, Argentina.\\ Full author list: \href{https://www.auger.org/archive/authors_2022_10.html}{https://www.auger.org/archive/authors\_2022\_10.html}
}

\abstract{%
  We test for an anisotropy in the mass of arriving cosmic-ray primaries associated with the galactic plane. The sensitivity to primary mass is obtained through the depth of shower maximum, \xmax{}, extracted from hybrid events measured over a 14-year period at the Pierre Auger Observatory. The sky is split into distinct on- and off-plane regions using the galactic latitude of each arriving cosmic ray to form two distributions of \xmax{}, which are compared using an Anderson-Darling 2-samples test. A scan over roughly half of the data is used to select a lower threshold energy of $10^{18.7}$\,eV and a galactic latitude splitting at $|b| = 30^\circ$, which are set as a prescription for the remaining data. With these thresholds, the distribution of \xmax{} from the on-plane region is found to have a $9.1 \pm 1.6^{+2.1}_{-2.2}$\,\gcm{} shallower mean and a $5.9\pm2.1^{+3.5}_{-2.5}$\,\gcm{} narrower width than that of the off-plane region and is observed in all telescope sites independently. These differences indicate that the mean mass of primary particles arriving from the on-plane region is greater than that of those from the off-plane region. Monte Carlo studies yield a $5.9\times10^{-6}$ random chance probability for the result in the independent data, lowering to a $6.0\times10^{-7}$ post-penalization random chance probability when the scanned data is included. Accounting for systematic uncertainties leads to an indication for anisotropy in mass composition above $10^{18.7}$\,eV with a $3.3\,\sigma$ significance. Furthermore, the result has been newly tested using additional FD data recovered from the selection process. This test independently disfavors the on- and off-plane regions being uniform in composition at the $2.2\,\sigma$ level, which is in good agreement with the expected sensitivity of the dataset used for this test. 
}
\maketitle

\section{Introduction}
\vspace{-1mm}
The Pierre Auger Observatory is currently the largest observatory dedicated to studying cosmic rays with energies in the EeV range, so-called ultra-high-energy cosmic rays, \textit{UHECR}~\cite{PierreAuger:2015eyc}. To do this, it uses both an array of particle detectors on the Earth's surface, the Surface Detector, \textit{SD}~\cite{PierreAuger:2007kus}, and an array of fluorescence telescopes monitoring the atmosphere above the SD, the Fluorescence Detector, \textit{FD}~\cite{PierreAuger:2009esk}. The highest quality data set of the Observatory is made up of UHECR events which have simultaneously been measured by both the FD and SD, so-called \textit{hybrid events}. In a hybrid event reconstruction, the geometry of the shower axis is highly constrained by combining the triggered pixel geometry/timing from the FDs and the high confidence core location/timing provided by the SDs. This results in an angular resolution for the pointing direction of the shower axis of better than $0.5^\circ$~\cite{Bonifazi:2009ma}, and a resolution on the location of the shower core of $50$\,m~\cite{Mostafa:2006id}. 

The low uncertainty geometric reconstruction provided by the hybrid method allows the evolution of the intensity of UV fluorescence light measured by the FD to be inverted to model the \textit{shower profile}, which is the number of charged particles in the air shower as a function of the amount of matter it has traversed, the \text{slant depth}, $X$. From the shower profile, the slant depth at which the maximum development of the shower occurs, \xmax{}, can be extracted. \xmax{} is closely related to the mass of the primary cosmic ray which induced the air-shower, but is subject to large fluctuations meaning that it can not be used on a shower-by-shower basis to determine primary mass. However, if collected with sufficient statistics, the first and second moments of distributions of \xmax{}, \xmaxmu{} and \xmaxsigma{} respectively, can be used to make high certainty estimations of the mean mass of the UHECR events used to form that particular \xmax{} distribution~\cite{Aab:2014kda}.

Up until recently~\cite{PierreAuger:2021jlg}, \xmax{} derived from hybrid measurements has been used to study the average composition of the cosmic-ray sky as whole, rather than being used to compare the mean compositions of different subsets of the sky. This choice was likely driven by the relatively sparse statistics available in hybrid studies due to the upper-limit of exposure available to them being set by the relatively low 14\,\% up-time of FDs.
The possibility of splitting the data set into subsamples was therefore limited by the need to maintain sufficient statistics to say something useful about primary composition. However, the Pierre Auger Observatory has now collected more than 14-years of FD data, and tens of thousands of high-quality hybrid measurements. With this quantity of data, the sky can be split into different regions and the mean mass of UHECRs arriving from them can be studied. This new reality then prompts two questions:
\begin{enumerate}
    \item Are the systematic uncertainties which trend with event arrival direction for hybrid reconstruction low enough to allow the mean mass arriving from different regions of the sky to be meaningfully compared?
    \item Is there a reason to expect that different regions of the sky may display differing compositions due to astrophysical causes?
\end{enumerate}

Question 1) will be explored in \autoref{sec:SystematicUncertanties}. For question 2) it is clear that the opportunity exists. The flux above the ankle at ${\sim}5$\,EeV~\cite{PierreAuger:2020kuy} is mixed in composition and has long been thought (now confirmed) to be extragalactic in origin~\cite{Linsley:1963bk}. Furthermore, it definitively displays anisotropy above 8\,EeV~\cite{Aab:2017tyv}. Additionally, as was nicely put by Alan Watson in 1990:
\begin{myquote}{3mm}
    \textit{``... the Larmor radius of a proton of $10^{18}$\,eV in a 3\,$\mu$G field is about 400\,pc, comparable to the thickness of the galactic disk. It follows, therefore, that, if the bulk of cosmic rays are protons, anisotropies associated with the magnetic field structure of the galactic disk might appear as the energy increases.''} - \cite{Watson:1990fj}
\end{myquote}
Indeed, there were hints of such a spectral feature starting somewhere around $10^{18.5}$\,eV for directions within $30^\circ$ of the galactic plane ~\cite{Szabelski:1986rx,Watson:1990fj}. Unfortunately, such an excess so far does not appear to be significant in the data of current experiments and therefore has not been given much attention since those initial publications. 

Now it is known that the flux above 1\,EeV is best described as an evolving mix of light-, intermediate-, and high-mass primaries~\cite{pierre2014aab, PierreAuger:2021mmt}. Due to the galactic magnetic field, \textit{GMF}, the different mass components present at any given energy will be deflected to different degrees as they travel from their extragalactic sources to Earth. This mass dependent deflection suggests that an anisotropy associated with the structure of the GMF would kick in for increasingly heavier components as energies climb. It is therefore distinctly possible that an anisotropy associated with the galactic plane could arise in the higher mass components of the flux at some energy in the EeV range.  

What follows below are specific tests for such a mass-dependent anisotropy associated with the galactic plane using hybrid data of the Pierre Auger Observatory collected between 2004 and the end of 2018. To avoid repeating the contents of the ICRC 2021 proceedings on this result~\cite{PierreAuger:2021jlg} in its entirety, the contents of this proceeding will aim to include components of the analysis that could not be fit in the eight pages allotted in that publication. Therefore, while this proceeding will cover the details of the analysis, increased space will be given to the cross checks and studies of the systematics uncertainties of the analysis. Additionally, a new test on an independent hybrid data set, recovered from the quality cuts, will also be discussed.
\vspace{-1mm}
\section{Reconstruction and selection}\label{sec:2}
\vspace{-1mm}
The reconstruction methods, selection cuts, and core analysis of \xmax{} distributions used here are largely the same as those described in~\cite{Yushkov:2020nhr}. A rigorous description of these methods can be found in~\cite{Aab:2014kda}. Other than the fiducial field-of-view cut, \textit{FidFoV}, which is treated below, a detailed description of the methods will not be provided here. 

The important differences from~\cite{Yushkov:2020nhr} are that the minimum energy for inclusion in the data set has been raised to $E > 10^{18.4}$\,eV, and that, as described below in \autoref{sec:ADEffects}, the event acceptance, reconstruction bias, and \xmax{} resolution are now treated based on the arrival direction of each event. The lower limit of $10^{18.4}$\,eV has been chosen as above this energy, the composition is well mixed and expected to be primarily of extragalactic origin \cite{Aab:2016zth}. The period over which data has been collected has also been slightly expanded to span events observed between 2004-12-01 and 2018-12-31, yielding 7572 high-quality events. A further subdivision of the data is necessary to test for the hypothesized anisotropy. Following the results of the scan shown in \autoref{fig:PRDScan}, the data is split into the events with a galactic latitude, $\abs{\,b\,} \leq 30^\circ$, the \textit{on-plane} sample, and, $\abs{\,b\,} > 30^\circ$, the \textit{off-plane} sample.

\begin{figure}[!htb]
    \centering
    \includegraphics[width=\columnwidth]{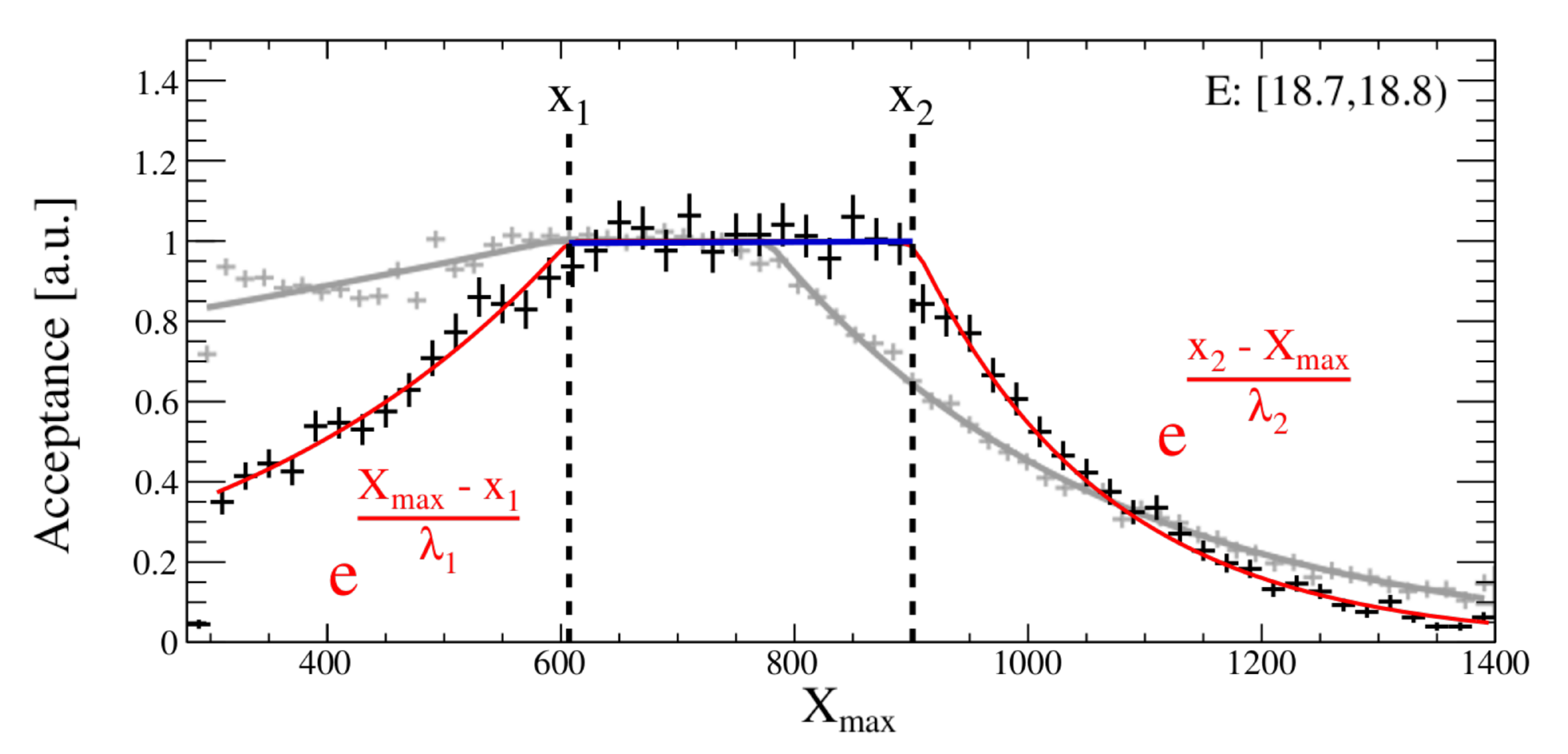}
    \caption{FD \xmax{} acceptance before (gray) and after (black) FidFoV cuts. A 4-variable parameterization of the post-FidFoV acceptance is shown in red. The range of \xmax{} values with unbiased sampling is shown in blue.}
    \label{fig:FidAcc}
\end{figure}

\vspace{-1mm}
\paragraph{Fiducial field-of-view selection}
\vspace{-1mm}
To ensure a good reconstruction of \xmax{} with the FD, \xmax{} itself should be directly observed. The telescopes of the FD have a field-of-view, \textit{FoV}, which is vertically constrained. Additionally, some showers in more vertical geometries will not reach \xmax{} before impacting the ground. These factors together lead to a geometric and \xmax{} dependence for what events end up in the analyzed data set. If unaccounted for, this \textit{\xmax{} acceptance}, will inevitably bias a composition study based on FD \xmax{} data. At the Observatory, this \xmax{} acceptance is addressed primarily through mitigation of the effect with the fiducial field-of-view, \textit{FidFoV}, cut. The FidFoV cut constrains the FD detector volume to only event geometries where the expected range of \xmax{} values would be visible in the FD FoV. As can be seen in \autoref{fig:FidAcc}, this changes the natural FD \xmax{} acceptance (gray) to one which is unbiased from ${\sim}600$ to ${\sim}900$\,\gcm{}, which spans the typical range of observed event \xmax{} values. Remaining effects on rare events outside this range are corrected using parameterizations as a function of \xmax{} and primary energy.

\subsection{Distributions of \xmaxbf{} and arrival direction}\label{sec:ADEffects}
\begin{figure}[!htb]
    \centering
    \vspace{2mm}
    \includegraphics[width=0.47\columnwidth]{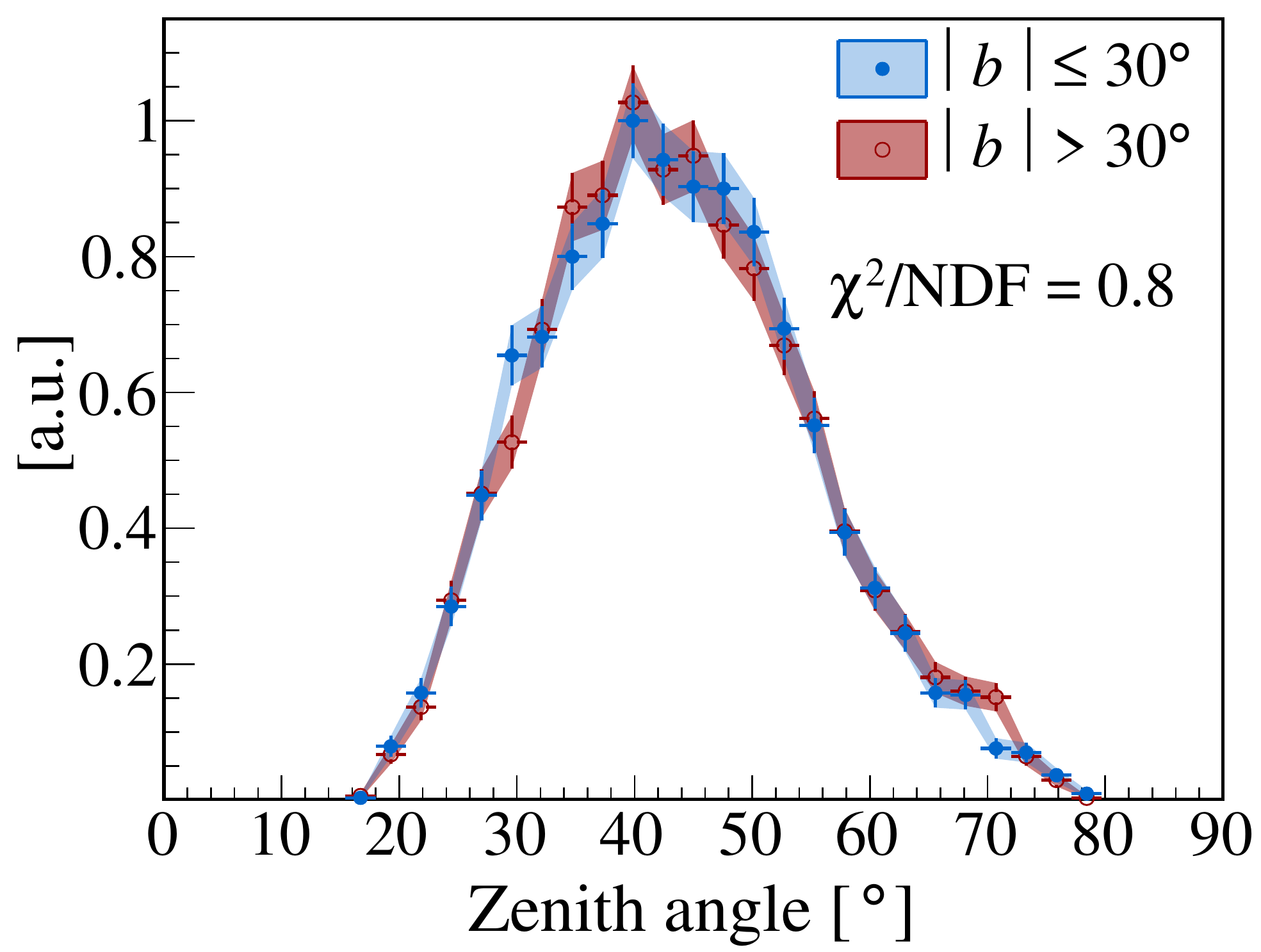}
    \includegraphics[width=0.47\columnwidth]{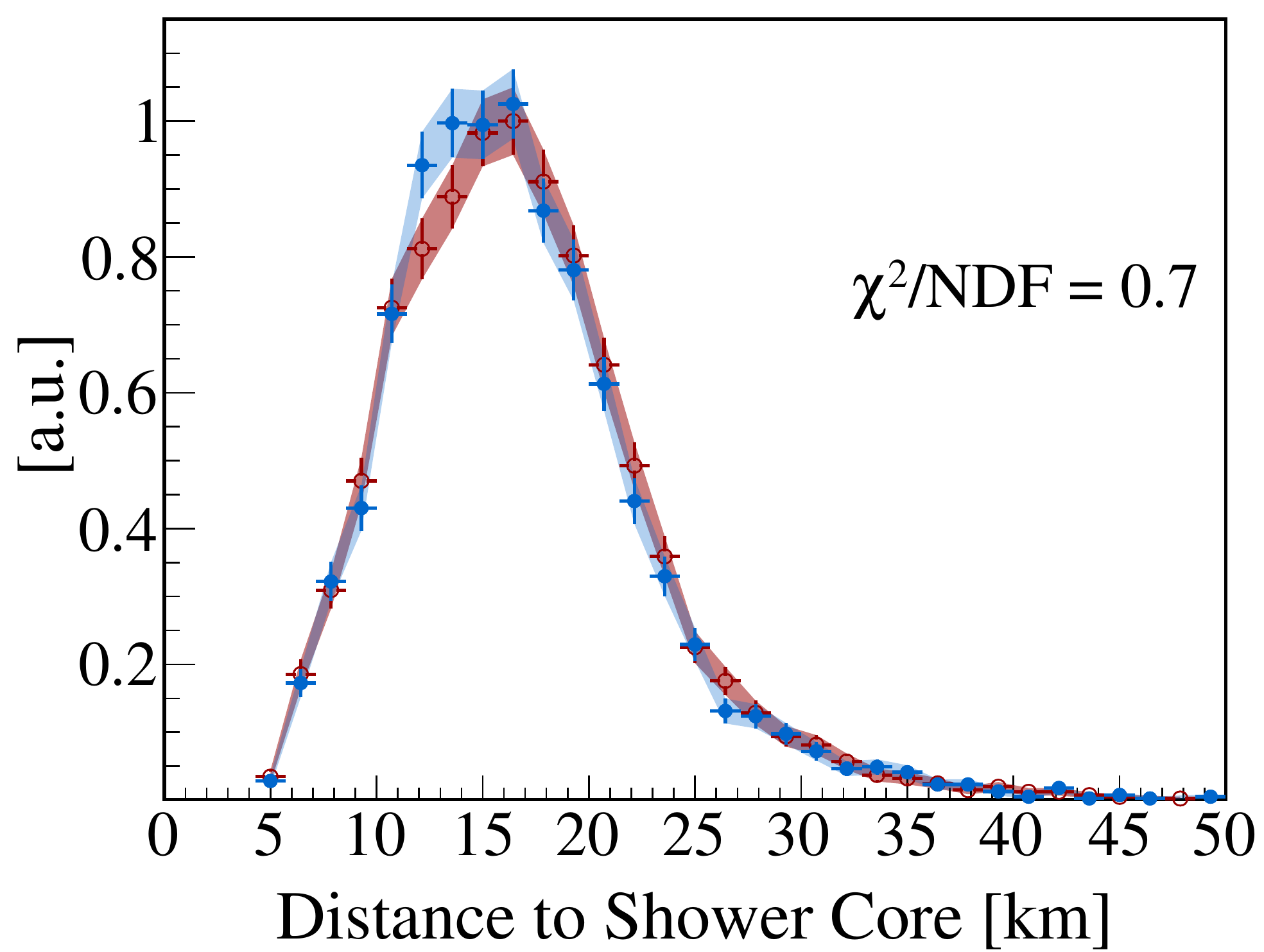}\\
    \includegraphics[width=0.47\columnwidth]{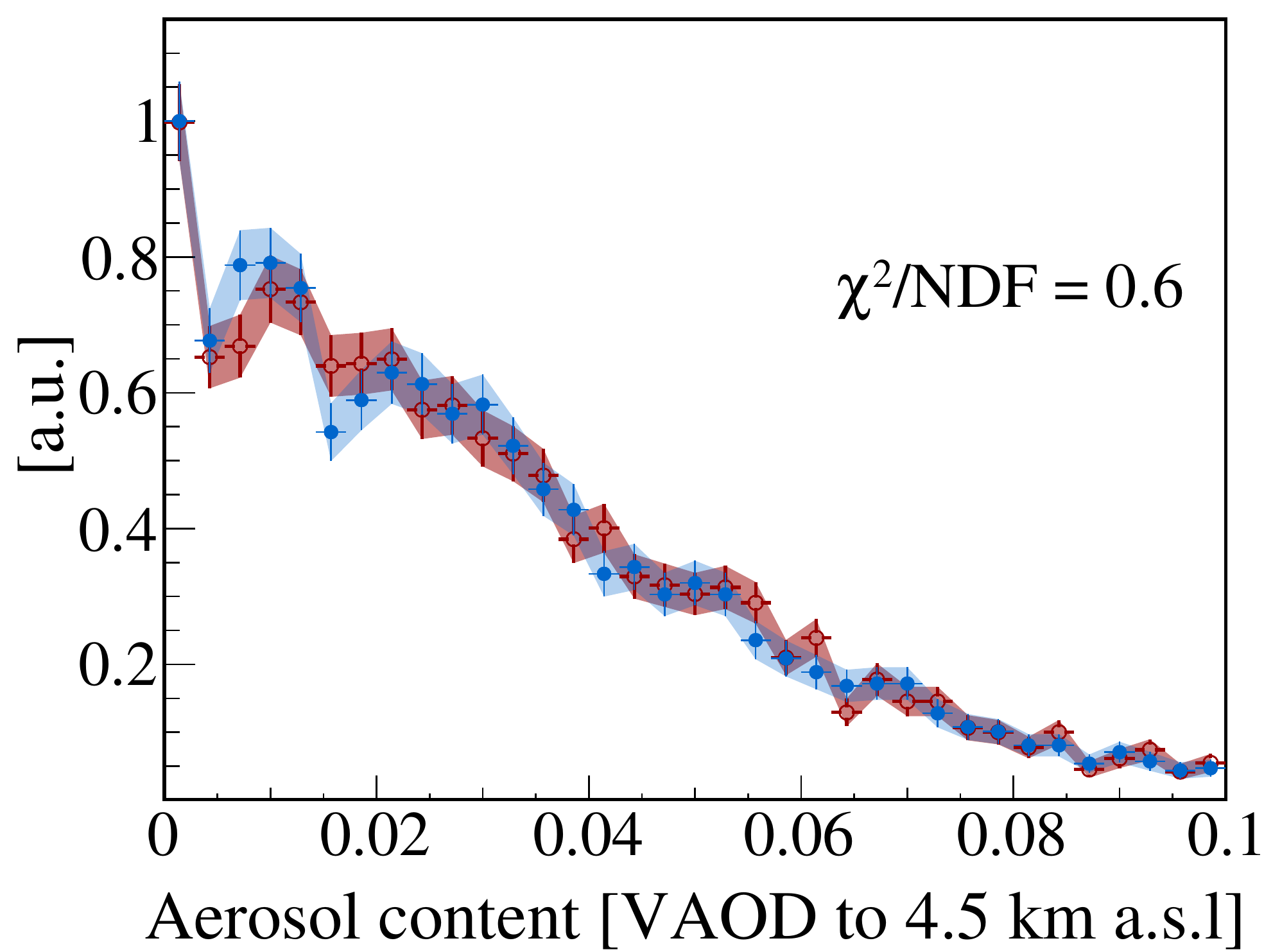}
    \includegraphics[width=0.47\columnwidth]{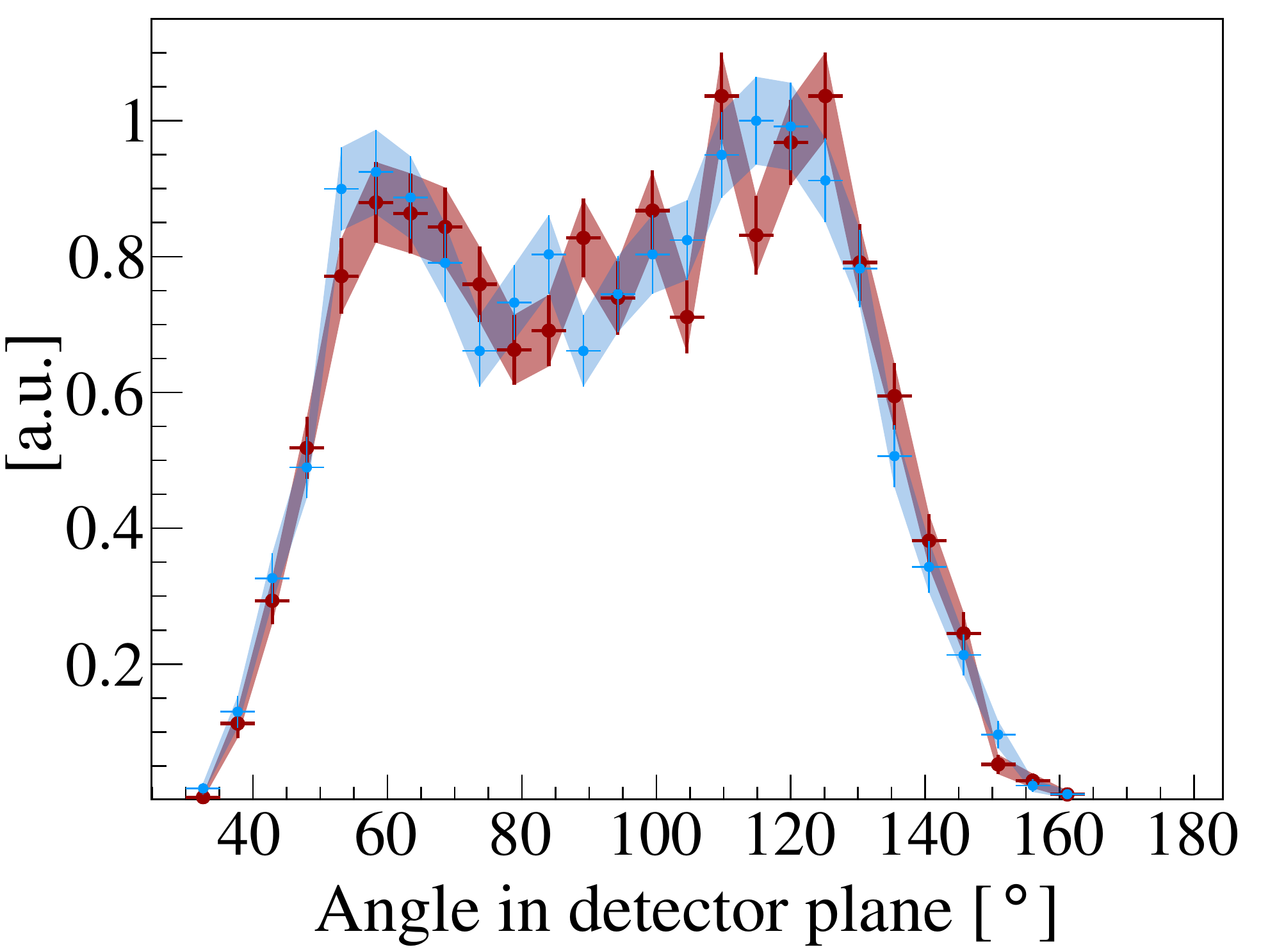}
    \vspace{-1mm}
    \caption{Geometries of the on- and off-plane samples.}\label{fig:Geometries}
    \vspace{3mm}
\end{figure}

After measurement, reconstruction, and selection, the observed \xmax{} distribution does not quite represent the true \xmax{} distribution of all cosmic rays landing within the Observatory. This is due to energy dependent biases on the reconstruction of \xmax{} ($B$), the resolution on \xmax{} of the hybrid reconstruction method ($R$), and the residual effects of the \xmax{} acceptance ($A$). Since the location of an event and its inclination with respect to the observing fluorescence telescope plays a role in the magnitude of $A$, $R$, and $B$, they have an inherent geometric dependence. If unaccounted for, this could bias this study. Distributions of the key geometric relationships between events and the FD can be seen in \autoref{fig:Geometries} for on- and off-plane regions. It is clear that there is little difference between the two regions, so $A$, $R$, and $B$ are expected to also be similar. 

To explicitly verify $A$, $R$, and $B$ similarity on and off the galactic plane, CONEX \cite{Bergmann:2006yz} is used to generate showers with Sibyll-2.3c~\cite{Riehn:2015oba}. The showers are then isotropically thrown into detector simulations which include the time-dependent state of the FD and SD from 2004 through 2018. These simulations, therefore, mimic the measurement conditions, trigger efficiency, and up-time of the real data, accurately modelling the exposure and geometries of events arriving from all parts of the sky~\cite{Abreu:2010aa}. Two sets of these simulations are produced, one formed from showers generated with a flat sampling of \xmax{} between 300 and 1500\,\gcm{}, the \textit{flat-MC}, and one formed from an equal number of proton, helium, nitrogen, and iron simulations which are then weighted to their abundances observed in data as reported in~\cite{Bellido:2017cgf}, the \textit{mixed-MC}. These simulated event sets are then subjected to the same reconstruction and selection techniques used on the real data, so that $A$, $R$, and $B$ are accurately included in them.

\begin{figure}
    \includegraphics[width=\columnwidth]{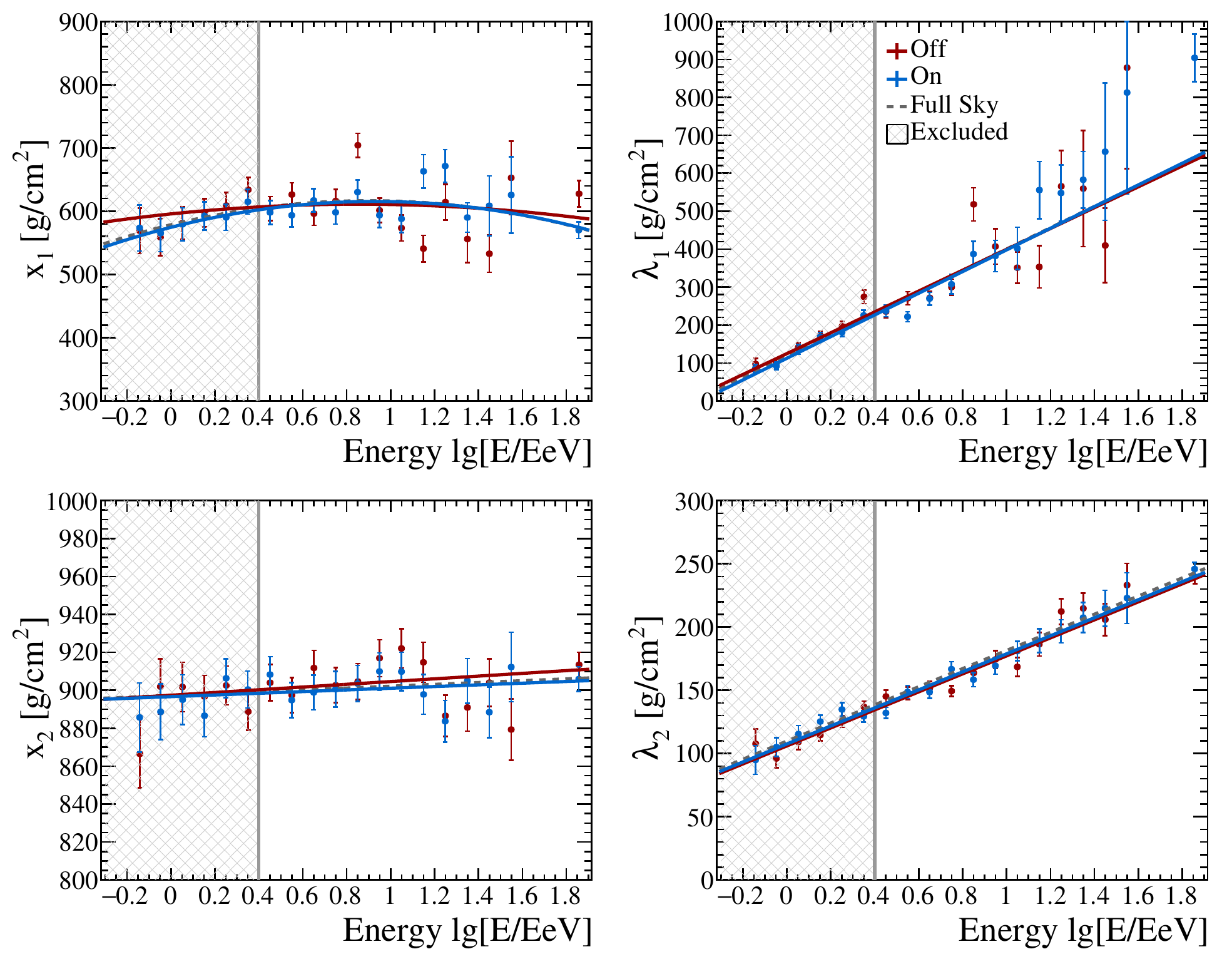}
    \vspace{-6mm}
    \caption{The \xmax{} acceptance parameterizations for the on- and off-plane sky regions from Monte Carlo.}
    \label{fig:OnOffAccParam}
\end{figure}

The flat-MC is split into on- and off-plane subsamples which are then used to extract the functional form of $A$ in 0.1\,\lge{} energy bins using the method illustrated in \autoref{fig:FidAcc}. The form of $A$ is extracted by leveraging the flatly sampled \xmax{} generation of the Monte Carlo, as, once the plateau of the distribution is normalized to one, the height of each bin represents the acceptance of events with \xmax{} values in that range. The acceptance is then fit with the 4-component parameterization illustrated in \autoref{fig:FidAcc}. The energy evolution of $x_1,\lambda_1,x_2$, and $\lambda_2$ is then parameterized separately with 2D polynomials for the on- and off-plane subsamples, resulting in \autoref{fig:OnOffAccParam}. From \autoref{fig:OnOffAccParam}, it is clear that above $10^{18.4}$\,eV there is no statistically significant difference in \xmax{} acceptance between the on- and off-plane regions. Even so, the region-specific parameterizations of $A$ are used to correct the 1.4\,\% of events with partial \xmax{} acceptance using the up-weighting technique outlined in~\cite{Aab:2014kda}. Uncertainties in these parameterizations result in the systematic uncertainties on the first and second moments specified in \autoref{tab:SysUncertainties}.

\begin{figure}
    \includegraphics[width=\columnwidth]{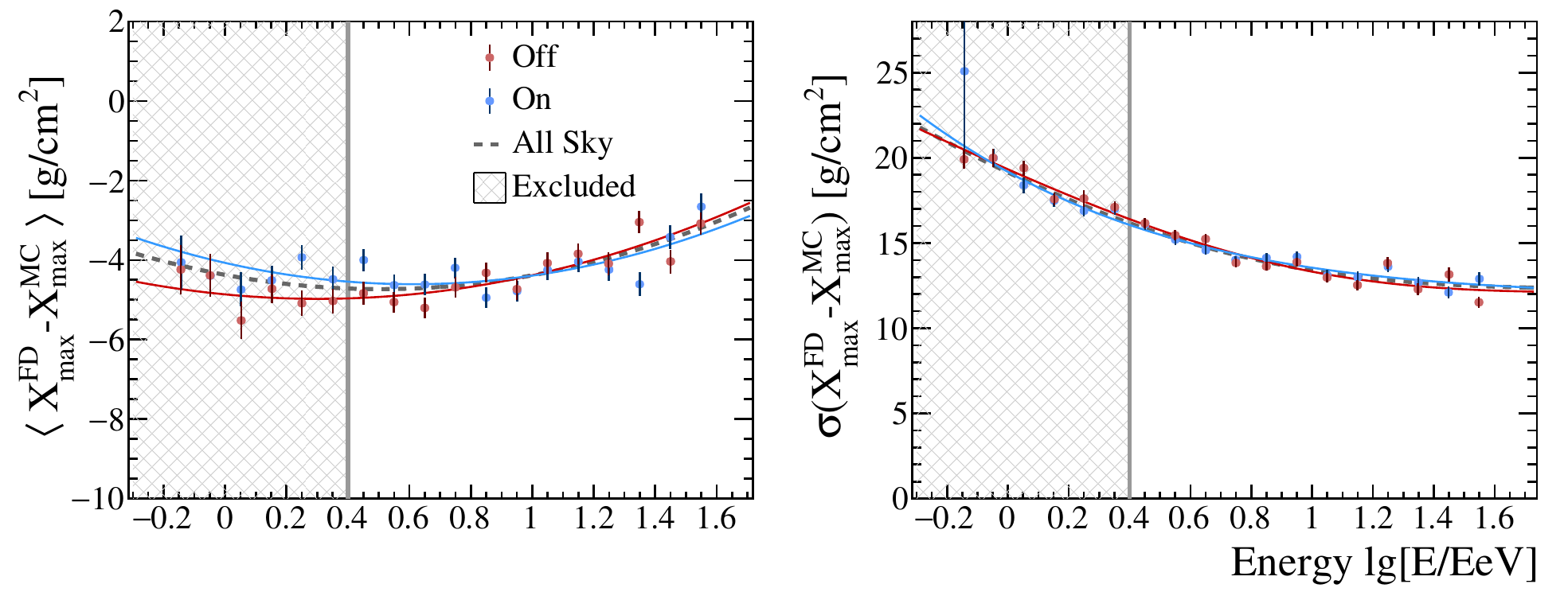}
    \vspace{-6mm}
    \caption{The \xmax{} reconstruction bias and resolution parameterizations for the on- and off-plane sky regions from Monte Carlo. Note: only the detectors and reconstruction \xmax{} resolution is shown. Other effects lowering the resolution are included as specified in~\cite{Aab:2014kda}.}
    \label{fig:OnOffBiasAndResParam}
\end{figure}

The mixed-MC is likewise split into on- and off-plane subsamples which are then used to extract the energy evolution of $B$ and $R$ for each region. $B$ and $R$ are extracted by forming a distribution of the difference between the FD reconstructed value of \xmax{} and the Monte Carlo truth value in 0.1\,\lge{} energy bins. From these, the mean reconstruction bias, $B=\langle X_{\rm max}^{\rm FD} - X_{\rm max}^{\rm MC} \rangle$, and the \xmax{} resolution, $R=\sigma \left(X_{\rm max}^{\rm FD} - X_{\rm max}^{\rm MC} \right)$, are extracted in each energy bin. Again, the evolution of each is parameterized with a 2D polynomial. \autoref{fig:OnOffBiasAndResParam} shows that $B$ and $R$ for the two regions are found to agree within errors. These however are also corrected for separately. Fit uncertainties again result in the systematic uncertainties outlined in \autoref{tab:SysUncertainties}. At this point, the \xmax{} distributions and moments from the on- and off-plane regions in each energy bin can be compared without bias from selection and reconstruction.

\section{Testing for anisotropy}\label{sec:TestingAnisotropy}
The specific hypothesis to be tested is whether, above some energy threshold, $E_{\rm th}$, the mean composition of UHECRs coming from directions near to the galactic plane is significantly higher in mass than those arriving further from it. This is to be tested using \xmax{} as a mass sensitive parameter. Typically, \xmax{} based composition analyses leverage the first two moments of \xmax{} distributions binned in energy, to comment on primary mass. This approach, however, does not lend itself well to quantifying the significance of a result testing the above statement. Instead, a test statistic, $TS$, which quantifies the degree of dissimilarity between the \xmax{} distributions in the two regions in a single value is preferred. For this, the returned value from the Anderson-Darling two-sample homogeneity test \cite{andersondarling}, \textit{AD-test}, has been selected as it scales with the dissimilarity of the tested distributions. The AD-test has good sensitivity to the full width of a distribution \cite{scholz1987k}, and has more power than the Kolmorogov-Smirnov test while remaining robust against false positives \cite{engmann2011comparing}.

To use the AD-test and \xmax{} for this purpose, two modifications are required. First, a single $TS$ comparing all events in each region above $E_{\rm th}$ is desired. So, all events with $E\geq E_{\rm th}$ in the on- and off-plane samples separately need to be collected into a common on-plane distribution and a common off-plane distribution. To do this, the natural evolution of \xmax{} with energy needs to be removed so that spectral features in the flux do not influence the result. Therefore, we define an energy-normalized \xmax{} value
\begin{equation}\label{eq:XmaxNorm}
X_{\text{max}}^{'} =  X_{\text{max}} -  \underbrace{\left(649 + 63.1 \, Z_{18} + 1.97 \, Z_{18}^{2}\right)}_\text{EPOS-LHC elongation rate for iron},
\end{equation}
where $Z_{18}=\log_{10}\left(E_\text{rec}/\,\text{EeV}\right)$. The last term in \autoref{eq:XmaxNorm} is the natural energy evolution of mean \xmax{} for iron primaries as predicted by EPOS-LHC~\cite{Pierog:2013ria}\footnote{Choice of hadronic interaction model varies result by $\sim0.02$\,\gcm{}.}. Second, the \xmaxnorm{} distribution of an on-plane sample populated with primaries which are on average heavier than those in the off-plane sample will display a lower mean and a narrower width than that of the off-plane \xmaxnorm{} distribution. Since the null hypothesis is that there is either no composition difference or a heavier off-plane sample, a $TS$ sensitive to the ordering of the \xmaxnorm{} distributions is required\footnote{Modifying the test to also require $\sigma( X_{\text{max}}^\prime)^{\rm on} < \sigma( X_{\text{max}}^\prime)^{\rm off}$ would be more restrictive, but conservatively has not been applied.}. The AD-test is insensitive to ordering, so it is modified to
\begin{equation}
TS =
\begin{cases}
    AD: \langle X_{\text{max}}^\prime \rangle^{\rm on} < \langle X_{\text{max}}^\prime \rangle^{\rm off} \\
    -3\hspace{1mm}: \text{else}
\end{cases},
\end{equation}
where $AD$ is the result of the AD-test comparing the on- and off-plane distributions, and $-3$ is selected as it is well below the minimum of the AD-test.

\vspace{-.1cm}
\myparagraph{Scan for energy and galactic latitude thresholds}
\vspace{-.1cm}
A scan has been used to select the optimal on/off splitting latitude, $b_{\rm split}$, and minimum energy, $E_{\rm th}$, as uncertainties in GMF models and source distributions make other approaches impractical. In this scan, each trial [$E_{\rm th}$, $b_{\rm split}$] pair is used to form on- and off-plane subsets and the $TS$ is extracted. To preserve the statistical strength of the sparse FD data set, a coarse scan of $5^\circ$ steps in $\abs{\,b\,}$ from $20^\circ$ to $35^\circ$ and 0.1\,\lge{} steps in energy from $18.4$ to $19.4$\,\lge{} is used. The scan is performed on the data set from~\cite{Aab:2014kda}, which includes events through Dec 31\textsuperscript{st} 2012. At the time of writing, this \textit{scan data set} represents $54\,\%$ of the analyzed events. The remaining $46\,\%$ of events, the \textit{post-scan data set}, is reserved as blind. 

\begin{figure}[!htb]
    \centering
    \includegraphics[width=0.45\textwidth]{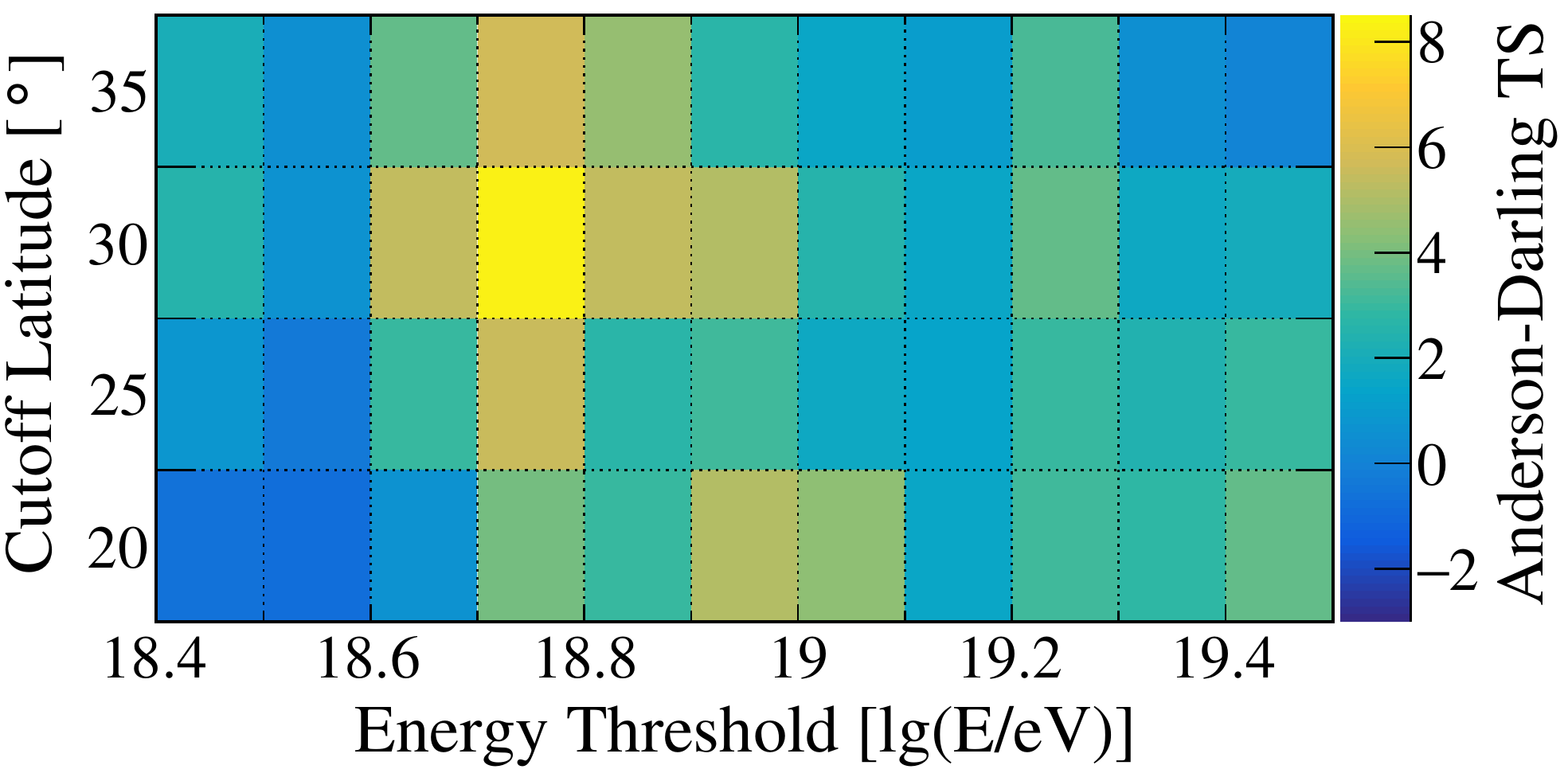}
    \vspace{-2mm}
    \caption{Parameter scan over 54\% of the data.}\label{fig:PRDScan}
\end{figure}

Interestingly, as shown in \autoref{fig:PRDScan}, all tested pairs result in $\langle X_{\text{max}}^\prime \rangle^{\rm on} < \langle X_{\text{max}}^\prime \rangle^{\rm off}$. An optimal [$E_{\rm th}$, $b_{\rm split}$] of [$10^{18.7}$\,eV,$30^\circ$] was found with a $TS = 8.4$. The selected [$E_{\rm th}$, $b_{\rm split}$] is applied as a prescription to the post-scan data set, which independently confirms the result with a $TS = 12.6$, for a total $TS=21.0$ for the full data set. 

\begin{figure}[!htb]
    \centering
    \includegraphics[width=.8\columnwidth]{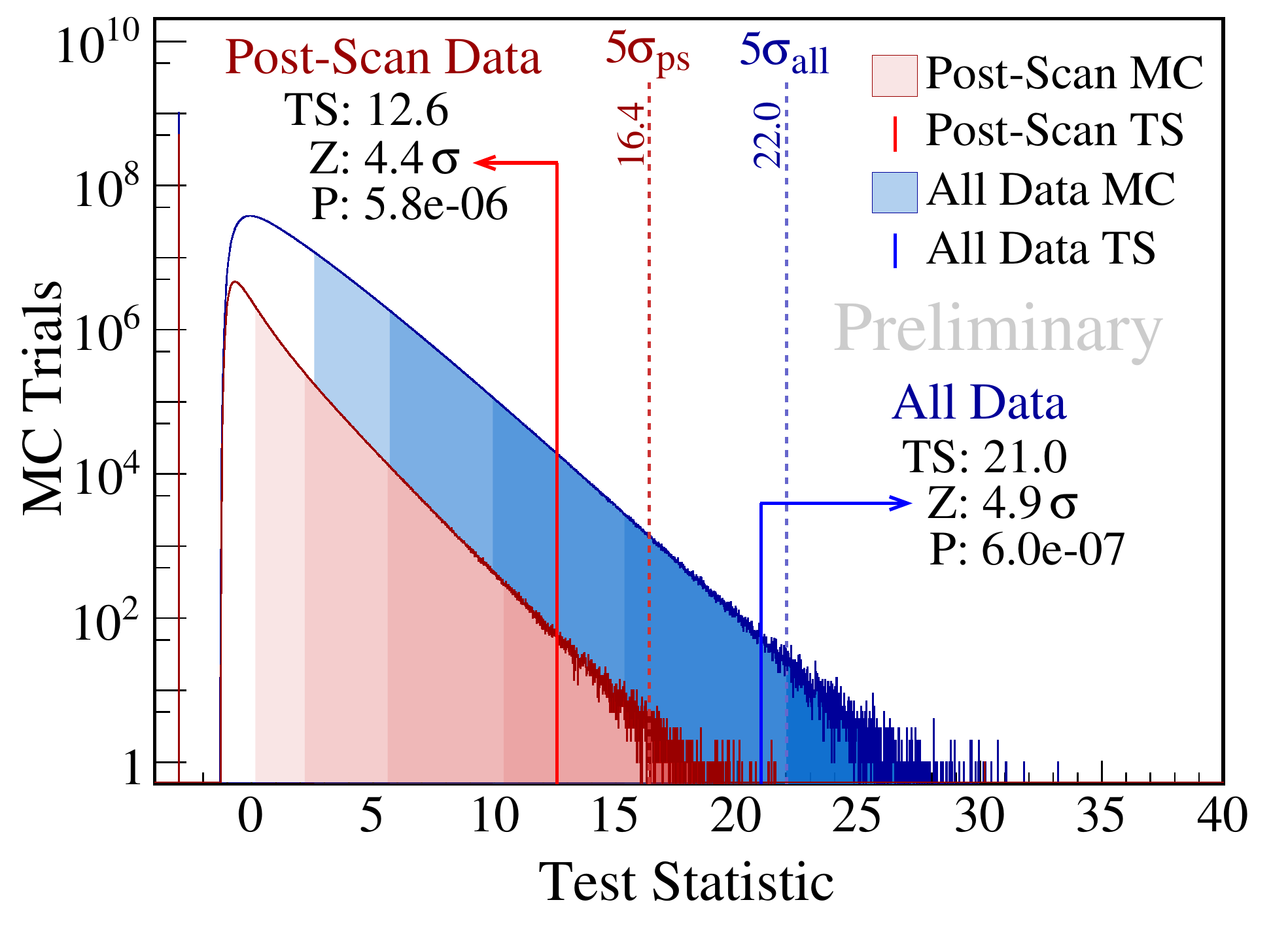}
    \caption{The Monte Carlo determination of the post-scan (red) and all-data (blue) significance with 1 and 10 billion randomized skies, respectively.}\label{fig:TStoSignificanceConversionNew}
\end{figure}

\vspace{-.1cm}
\myparagraph{Statistical significance}
\vspace{-.1cm}
The chance probability of the observed TS occurring with in an isotropic sky is tested using Monte Carlo methods on randomized skies derived from the real data. To form each randomized sky, the arrival direction is first decoupled from the energy and \xmaxnorm{} values of each event. These are then randomly re-paired to create a new sky which maintains the real \xmax{}, energy, and sky exposure distributions, but has a scrambled arrival direction/composition pairing. The above analysis is then used to extract a $TS$ from each sky which is compared to the result in data. Skies which display more extreme on-/off-plane differences than those observed in data are tallied and used to calculate the probability of an isotropic sky generating the observed $TS$. The results of this procedure are shown in  \autoref{fig:TStoSignificanceConversionNew}.

For the blind, post-scan data set, the prescribed [$E_{\rm th},b_{\rm split}$] pair is used to split each randomized sky into on- and off-plane samples and a $TS$ is extracted. In one billion random skies, only 5865 resulted in a more extreme $TS$ than the 12.6 observed in data. This indicates a chance probability of $5.87\times10^{-6}$ which corresponds to 4.4\,$\sigma$. 

To calculate the significance of the result when the scan- and post-scan data sets are combined, the entire analysis chain, including the scan, is duplicated. In each random sky, 54\,\% of the data is used to scan for the [$E_{\rm th},b_{\rm split}$] pair which results in the most extreme result, fully penalizing for the scan. These values are then used to split all data in the random sky into on- and off-plane subsamples and the $TS$ for the sky is extracted. From 10 billion random skies, only 5964 resulted in a more extreme $TS$ than the 21.0 observed in data. This indicates to a chance probability of $5.96\times10^{-7}$ which corresponds to 4.9\,$\sigma$. The strong penalization of the scanned data is evident as the additional 54\,\% of the data (with \Dxmaxmunorm{} $= 8.5$\,\gcm{}) only resulted in an 11\,\% increase of the significance of the observation. 

\myparagraph{\xmax{} moments and trends}

\begin{figure*}[!hbt]
\centering
    \begin{minipage}{.63\textwidth}
      \centering
      \captionsetup{width=.9\linewidth}
      \includegraphics[width=.49\textwidth,valign=t]{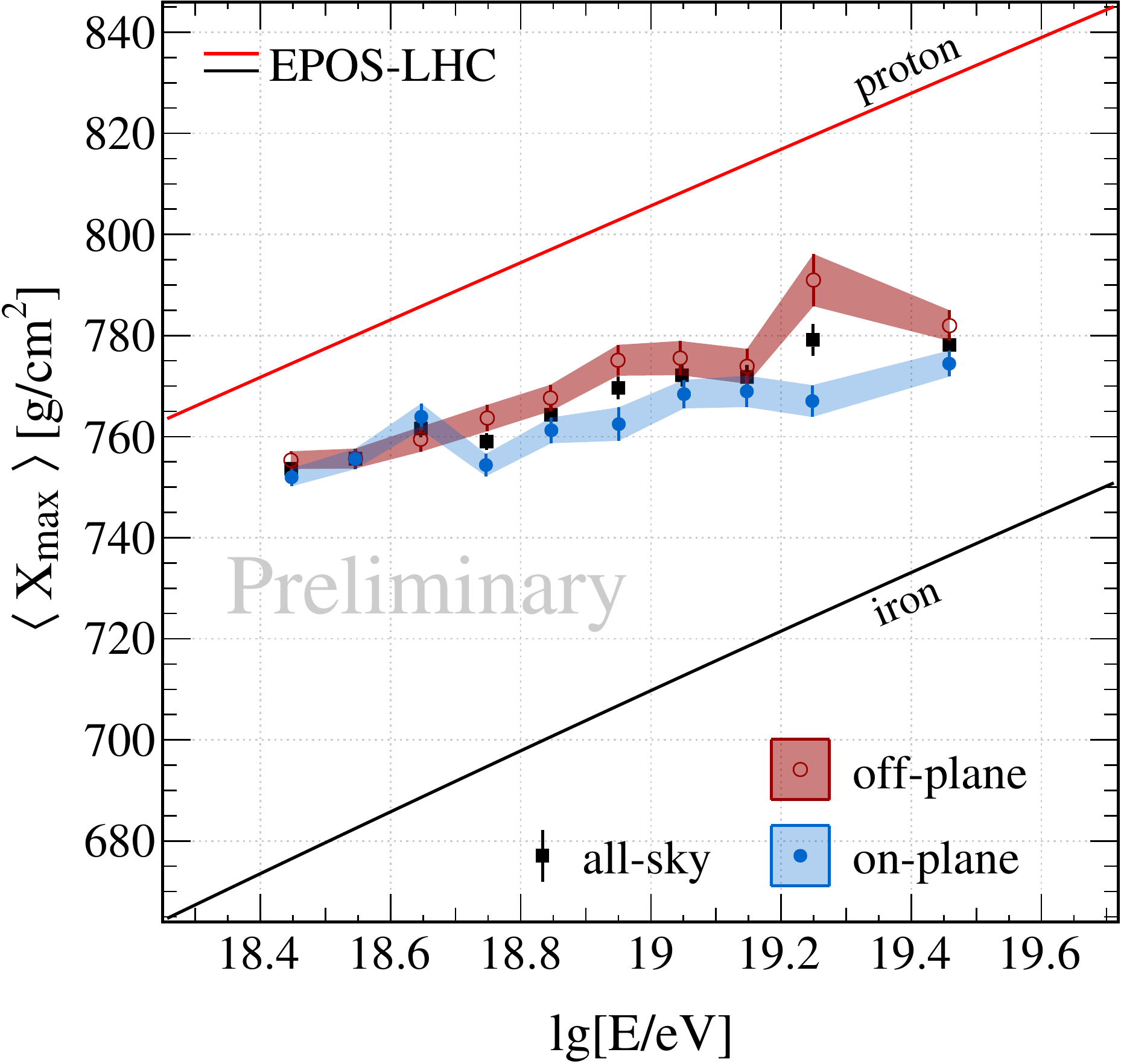}
      \includegraphics[width=.49\textwidth,valign=t]{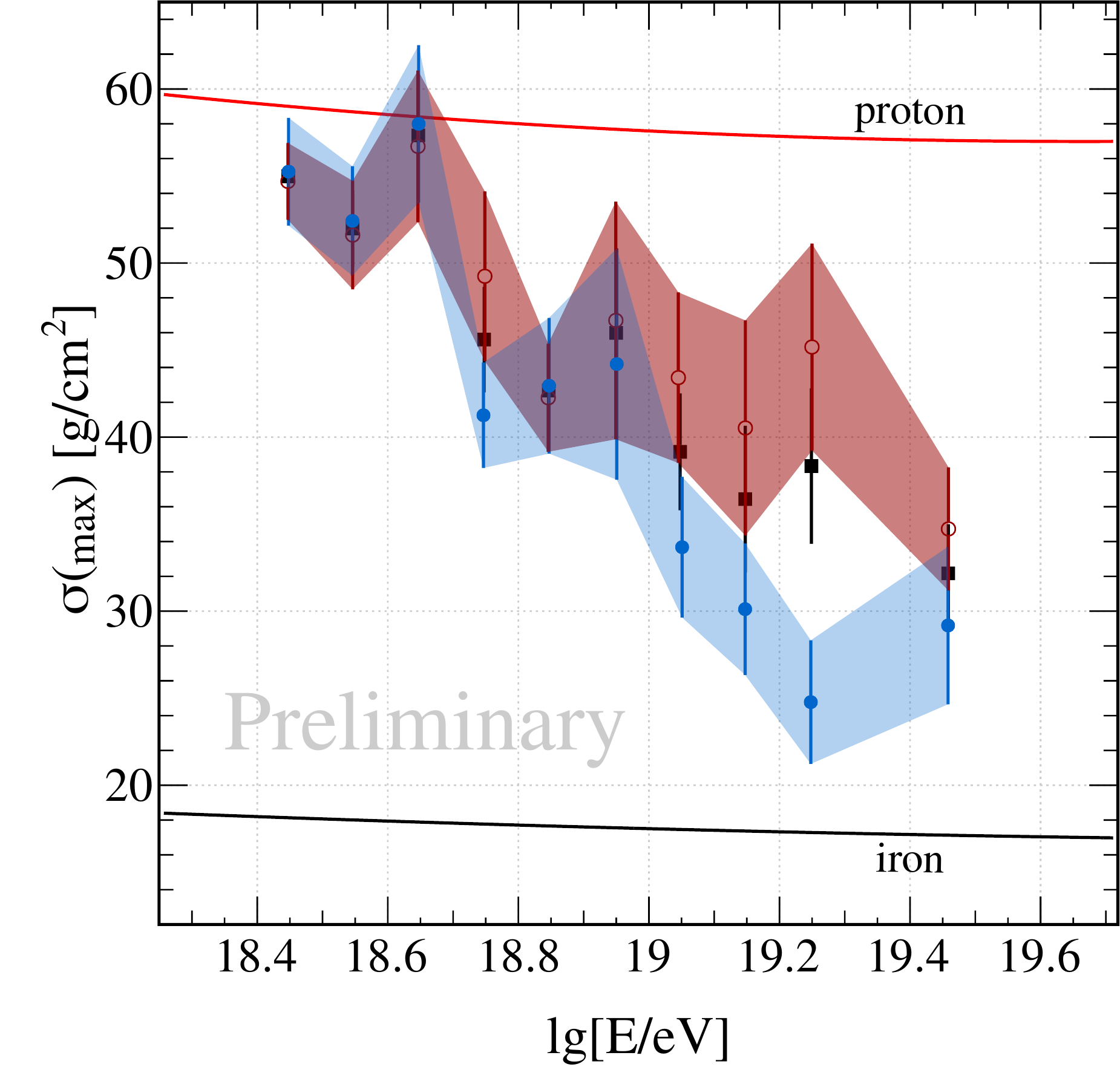}
      \vspace{-1mm}
      \caption{The first (left) and second (right) moments of the \xmax{} distributions from on- and off-plane regions.}
      \label{fig:CompositionPlots} 
    \end{minipage}%
    \hfill
    \begin{minipage}{.35\textwidth}
      \centering
      \vspace{-1mm}
      \includegraphics[width=\textwidth,valign=t]{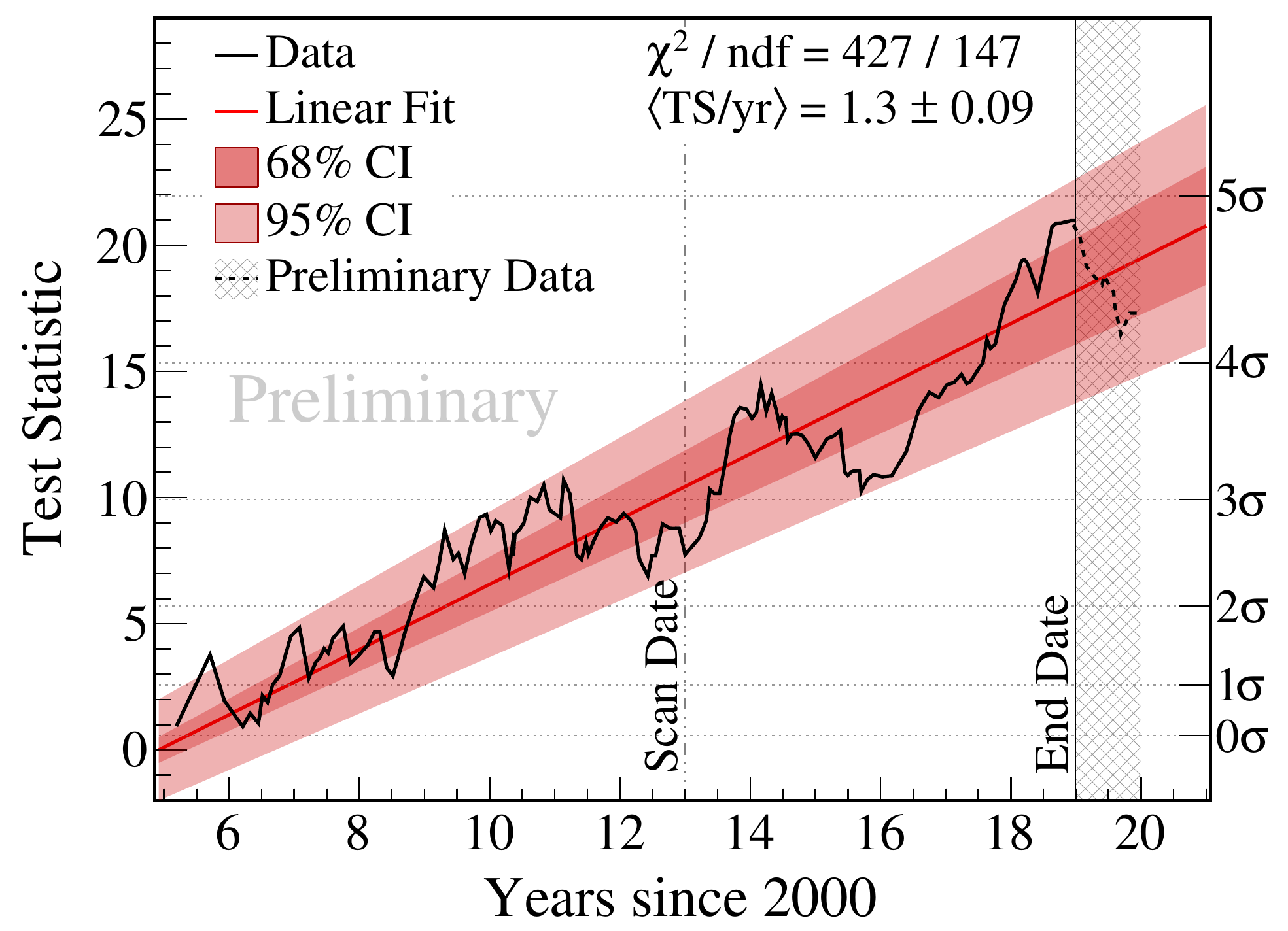}
      \vspace{1.5mm}
      \captionof{figure}{The time evolution of the TS with significance indicated on the right. The shaded region is preliminary data.}
      \label{fig:TimeEvolution}
    \end{minipage}
\end{figure*}

To illustrate the difference in composition on and off the plane, the first two moments of the \xmax{} distribution in each 0.1\,\lge{} energy bin has been plotted in \autoref{fig:CompositionPlots} for both regions. Above $10^{18.7}$\,eV there is a clear separation in \xmaxmu{} for all energy bins. Most energy bins also display a separation in \xmaxsigma{}. Heavier primaries are expected to, on average, have a shallower \xmax{} and lower shower-to-shower fluctuations. Therefore the correlated difference seen here indicates that, for this data sample, primaries from the on-plane region have a higher mean mass than that of the off-plane region above $10^{18.7}$\,eV.

To evaluate the degree to which fluctuation plays a role in the observed result, the growth of the $TS$ over time has been plotted in \autoref{fig:TimeEvolution}. The time evolution of the signal is consistent with linear growth at a rate of 1.3\,$TS$\,yr$^{-1}$. This behavior is in line with expectations for a real difference in mean mass between the subsamples. The shaded region of \autoref{fig:TimeEvolution} shows preliminary data from 2019. These reconstructions were not subject to a validated reconstruction chain and may change. Still, when added, a 3.7/4.4\,$\sigma$ (post-scan/all data) statistical significance is expected. The best fit rate of growth of 1.3\,$TS$\,yr$^{-1}$ remains unchanged.

\section{Systematic uncertainties}\label{sec:SystematicUncertanties}
\autoref{fig:Geometries} shows that there is very little difference between the on- and off-plane regions in the local reference frame of the detector. The data sets from the two regions also consist of events measured with the same instrumentation in the same location, and have been reconstructed with the same methods. Because systematic uncertainties are derived from measurement effects in the local frame, these similarities between the on- and off-plane samples mean that the majority of uncertainty sources outlined in~\cite{Aab:2014kda} will apply equally to both regions and thereby cancel in a comparison. Furthermore, from the acceptance, resolution, and bias studies in \autoref{sec:ADEffects}, the two regions are also free from selection and reconstruction biases.

\begin{table}[!htb]
    \vspace{2mm}
    \centering
    \caption{Systematic uncertainties on the difference in the means, \Dxmaxmu{}, and widths, \Dxmaxsigma{}, of the on- and off-plane \xmax{} distributions.}\label{tab:SysUncertainties}
    \vspace{-1mm}
    \renewcommand*{\arraystretch}{1.1}
    \setlength{\tabcolsep}{0.5em}
        \begin{tabular}{l|cc}
            \multirow{2}{*}{Source} & \multicolumn{2}{c}{ Uncertainty [\gcm{}] of}\\
            & \Dxmaxmu{} & \Dxmaxsigma{}\\\hline\hline
            $A$ correction  & $^{+1.14}_{-0.71}$ & $^{+2.37}_{-1.61}$ \\
            $B$ correction  & \small{$\pm 0.36$} & \small{$\pm 0.01$} \\
            $R$ correction  & 0                  & $^{+1.78}_{-0.24}$ \\
            Seasonal        & $^{+1.00}_{-1.53}$ & $^{+1.19}_{-1.23}$ \\
            Instrumentation & \small{$\pm 1.41 $}& \small{$\pm 1.41 $}\\ \hline
            Sum in Quadrature         & $^{+2.10}_\mathbf{-2.23}$ & $^{+3.49}_\mathbf{-2.48}$ \\
        \end{tabular}
    \vspace{-1mm}
\end{table} 

\begin{figure*}[!htb]
            \includegraphics[width=\columnwidth]{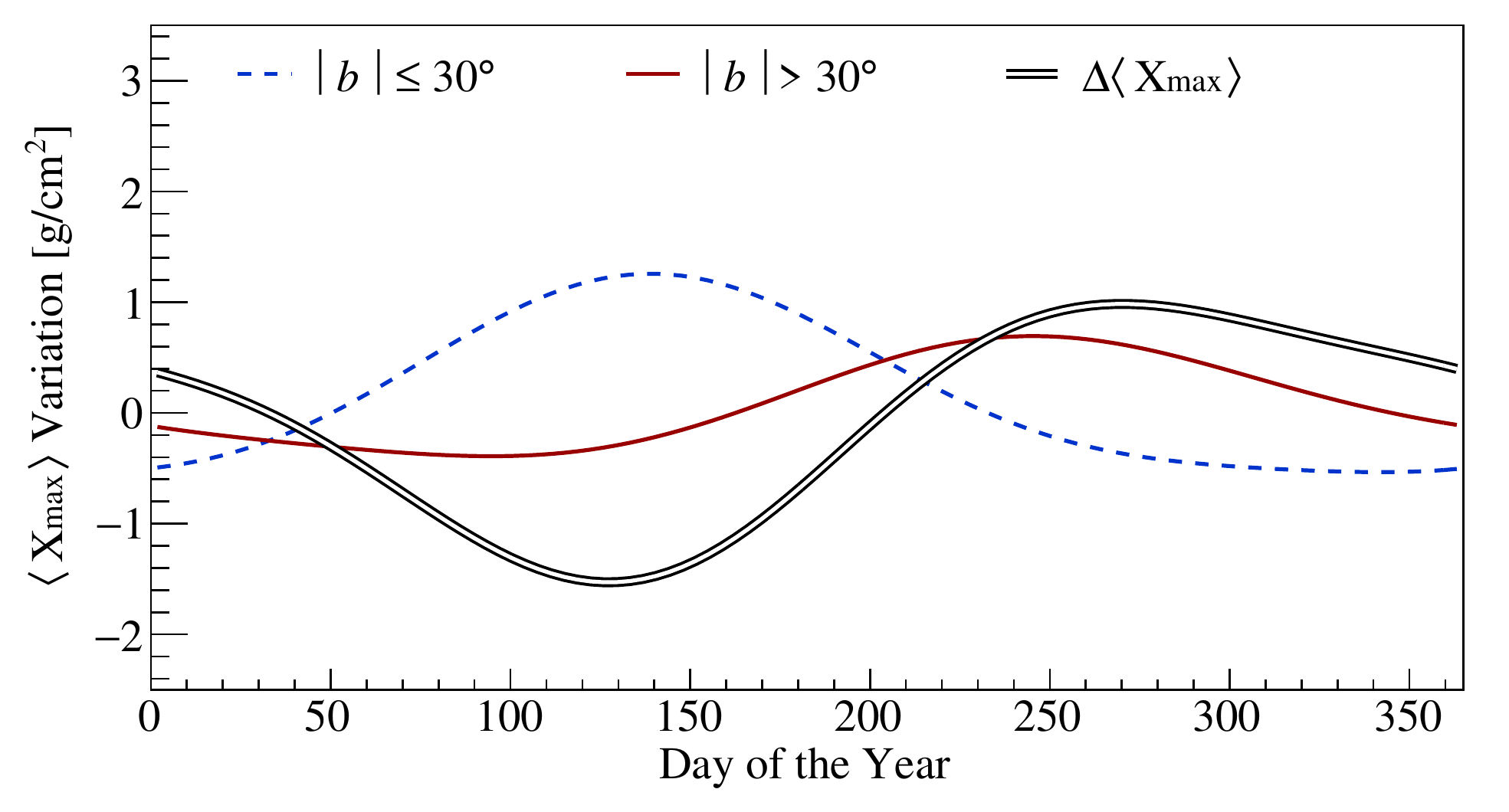}\hspace{2mm}
            \includegraphics[width=\columnwidth]{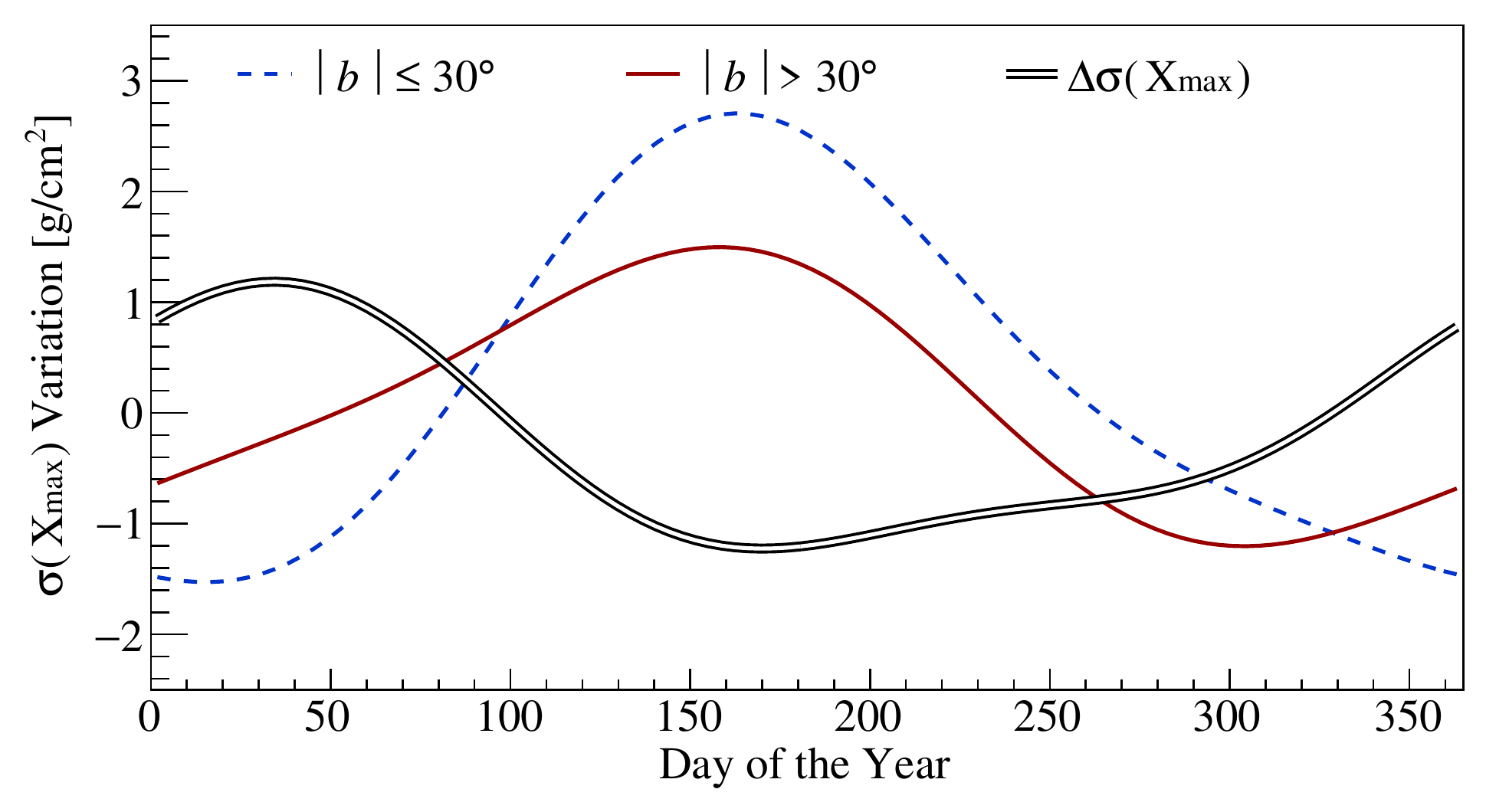}
            \caption{Seasonal fluctuation of the first (left) and second (right) \xmax{} moments for the on-plane (blue) and off-plane (red) samples. The  double black line shows the difference between these curves. The maximum and minimum values of these differences are taken as the systematic uncertainty on \Dxmaxmunorm{} and \Dxmaxsigmanorm{}.}\label{fig:seasonal}
\end{figure*}

To test for potential systematic effects derived from uncertainties in the $A$, $B$, and $R$ corrections, possible seasonal effects, and small differences between the instrumentation at different fluorescence telescope sites, \textit{FD-sites}, several studies were performed. All permutations of the uncertainties in the $A$, $B$, and $R$ corrections were evaluated and the maximum changes in \Dxmaxmunorm{} and \Dxmaxsigmanorm{} were recorded. To look for on-/off-plane biases associated with instrumentation differences, events seen by two or more FD-sites were used to compare the \xmax{} reconstructions of each site. No significant biases were found, and the instrumentation-derived systematic uncertainties from~\cite{Aab:2014kda} were adopted. To check for systematics derived from a combination of the seasonal dependence of the FD exposure and normal yearly variation of the atmospheric quality, \xmaxmunorm{} and \xmaxsigmanorm{} for the on- and off-plane regions were tracked over the course of the year, and is shown in \autoref{fig:seasonal}. The maximum exposure-weighted difference between the two regions at any point of the year has been adopted as the systematic uncertainty due to seasonal effects. These uncertainties and the total systematic uncertainty on \Dxmaxmunorm{} and \Dxmaxsigmanorm{} are listed in \autoref{tab:SysUncertainties}.


\myparagraph{Confidence level considering systematic uncertainties}
The observed \Dxmaxmunorm{} of $9.1\pm1.6$\,\gcm{} is 4.1 times larger than the 2.2\,\gcm{} systematic uncertainty listed in Table \ref{tab:SysUncertainties}. The observed \Dxmaxsigmanorm{} of $5.9\pm2.9$\,\gcm{} is 2.4 times larger than its 2.5\,\gcm{} systematic uncertainty. This means it is unlikely that the result could be entirely due to systematic effects. However, the systematic uncertainties in \Dxmaxmunorm{} and \Dxmaxsigmanorm{} may increase the likelihood of an extreme result occurring in data. To quantify the result significance taking possible systematic effects into account, a two step approach is taken. First, the on-/off-plane difference is reduced by adding a value sampled from a Gaussian distribution with $\mu = 2.2$\,\gcm{} and $\sigma = 2.5$\,\gcm{} to the on-plane sample. Then, the AD-test is applied to the resulting on- and off-plane distributions. Repeating this process 1 million times results in a mean TS of $11.3\pm0.5$
\footnote{Treating the other side of the systematic errors in the same way results in a $TS$ of $31.8\pm1.1$ ($6.3\,\sigma$).}. 
If these values are converted to significances using the data from \autoref{fig:TStoSignificanceConversionNew}, this corresponds to at least 3.3\,$\sigma{}$. Conservatively, to include systematic effects, this lower bound of 3.3\,$\sigma{}$ is adopted as the confidence level of the result.

\begin{figure}[!htb]
    \centering
    \includegraphics[width=.45\textwidth]{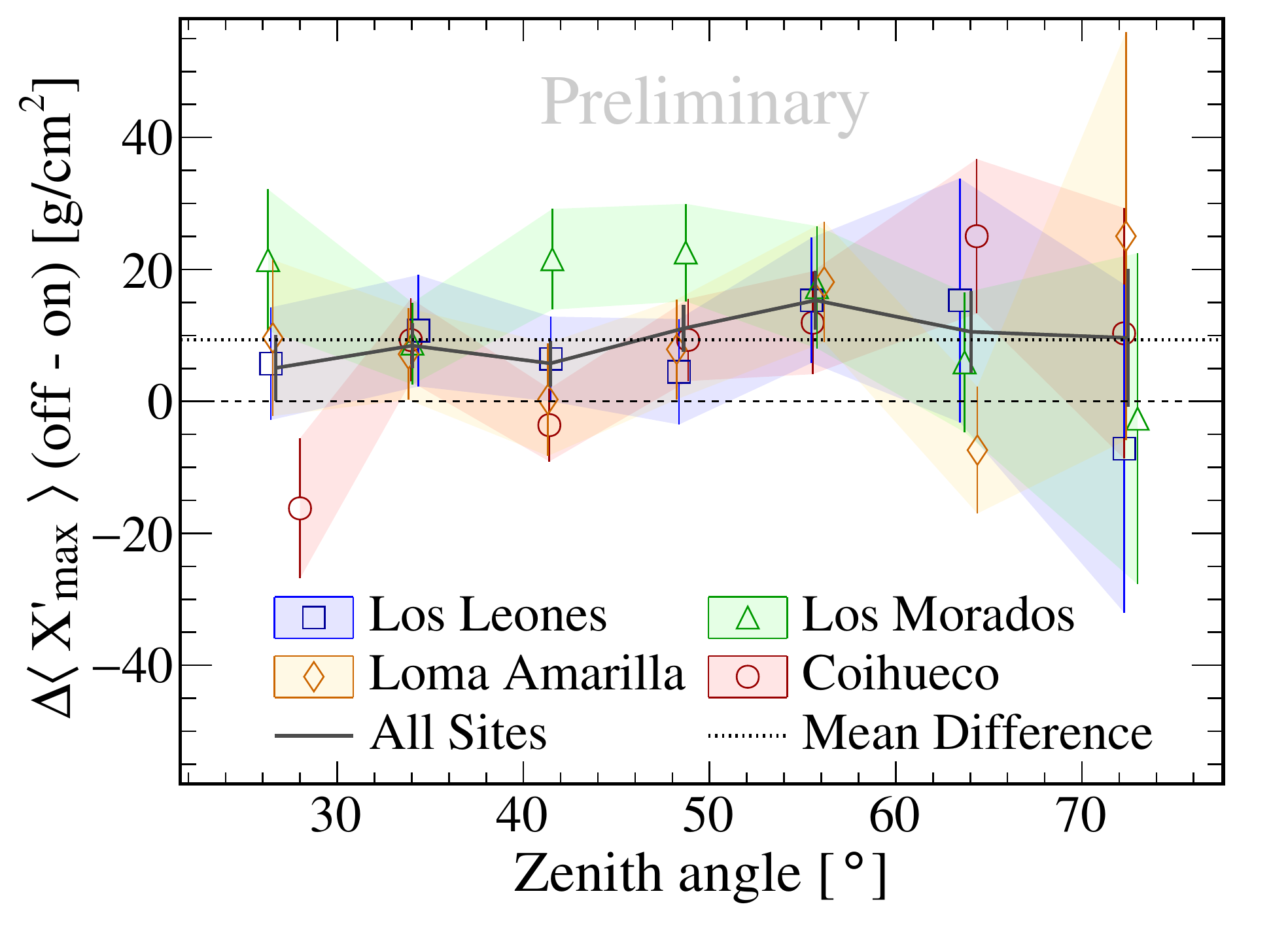}
    \vspace{-3mm}
    \caption{\Dxmaxmunorm{} by FD-site and zenith bin.}\label{fig:EyeZenith}
\end{figure}

\vspace{-.1cm}
\myparagraph{Cross-check: Results by zenith angle and FD-site}
If astrophysical in nature, the difference in composition of UHECR arriving from the on- and off-plane regions should be independently observed by each FD-site and in all zenith angle ($\theta$) ranges. \autoref{fig:EyeZenith} shows that the difference in \xmax{} on and off the plane is indeed present in all zenith angles bins and is also observed by all FD-sites independently. Even more stringently, when the response of each FD-site is split in $\cos^2\theta$ bins, it appears in 22 out of  28 bins. This independent observation at all sites and zeniths is a strong confirmation of the stereo study described above, showing that detector systematics can not play a large role in the result. Furthermore, because the FD-sites have FoVs differing by $90^\circ$ on average, each sees the galactic plane at a different local geometry and time during the year, making it unlikely that some unidentified detector, reconstruction, or atmospheric effect is causing the observed anisotropy. 

\section{Independent Test of the Result}

To cross-check the validity of the previous results, an independent data set is needed. The selection presented in \autoref{sec:2} aims to select only high-quality events. In particular, the \textit{FidFoV} cut is needed to ensure that the distribution of selected events covers the expected range of \xmax{} values with an unbiased acceptance. The events rejected by the FidFoV cut are still high-quality and perfectly well reconstructed. In this section, a second data set is built from the events removed in this FidFoV cut to form an \textit{out-FidFoV} data set which is 82.9\,\% as large as the data set already used in this analysis, the \textit{in-FidFoV} data set.

\begin{figure}[!hbt] \centering
    \includegraphics[width=0.42\textwidth]{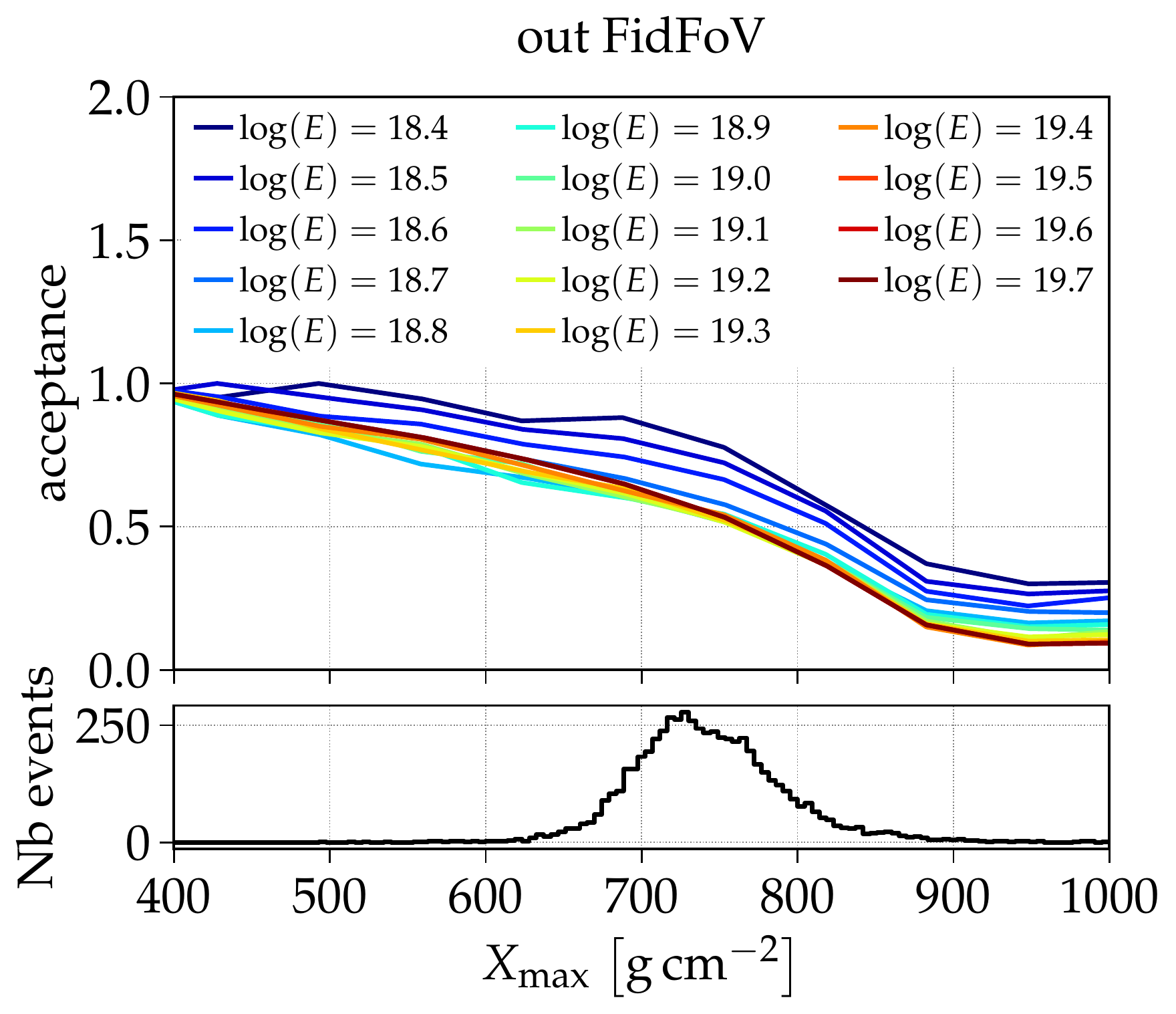}
    \vspace{-2mm}
    \caption{Top panel: acceptance for out-\textit{FidFoV} data set split in 0.1\,\lge{} energy bins. The energy in the legend is the lower energy edge of the bin, excepting 19.7 which corresponds to $\log(E/\text{eV}) \geq 19.7$. Lower panel: the \xmax{} distribution of events for the full energy range.}
    \label{fig:acceptance}
\end{figure}

\autoref{fig:acceptance} represents the \xmax{} acceptance of this out-FidFoV data set in different energy ranges.  For all energies, the efficiency maximized for low \xmax{} values and decreases as \xmax{} increases. Indeed, without the FidFoV selection, deep showers tend to be under-represented \citep{Aab:2014kda}. Thus, the new data set is made out of shallower \xmax{} events. As consequence, the \xmaxmu{} distribution of the new data set is on average $10$ g/cm$^2$ shallower. Despite this bias, if the difference between on- and off-plane is astrophysical, it should also appear in the out-FidFoV data set.

\begin{figure}[hbtp] \centering
    \includegraphics[width=0.4\textwidth]{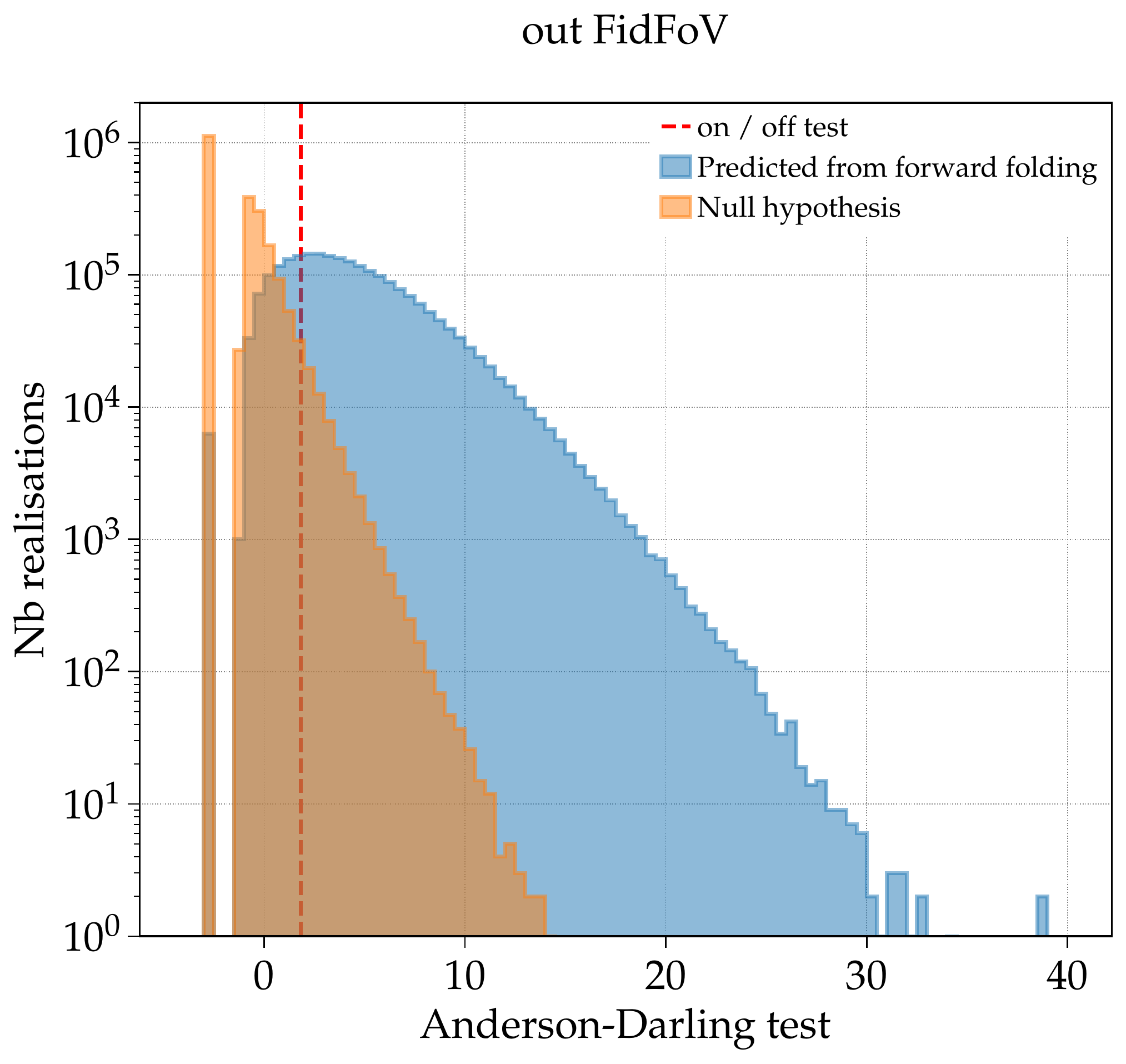}
    \vspace{-2mm}
    \caption{Anderson-Darling test from randomized skies (orange) and from forward folding (blue) for the out-\textit{FidFoV} data set. The red dashed line indicates the value obtained for data in the on/off test.}
    \label{fig:FF_TS}
\end{figure}

To test this hypothesis, this new data set is divided into on- and off-plane samples using the $E_{\rm th} = 10^{18.7}$\,eV, $b_{\rm split}=30^\circ$ splitting determined by the scan. The resulting on/off distributions show a somewhat smaller \Dxmaxmu{} of ${\sim}5$g/cm$^2$, which is only 55\,\% of what was obtained with the in-FidFoV data set. The AD-test returns a $TS = 1.8$. The post-trial significance of this test is re-evaluated with the method in \autoref{sec:TestingAnisotropy}, using two million randomized MC trials generated from the out-FidFoV data set. The corresponding $TS$ distribution is represented by the orange histogram in \autoref{fig:FF_TS}. The red dashed line depicts the AD-test obtained for the data set. The corresponding significance is ${\sim}2.2\sigma$, which gives a probability of 0.03 that this would result from an isotropic sky. The significance seen in this sample is, therefore, much lower than with the in-FidFoV data set. 

The lower significance of the out-FidFoV sample begs the question, why is there such a large difference between the in-FidFoV and out-FidFoV data sets? The bottom panel of \autoref{fig:acceptance} shows the distribution of \xmax{} values from events. Clearly the bulk of the distribution lies in a region where the acceptance is decreasing quickly. It is possible that this could reduce the sensitivity to a difference in composition. To test this hypothesis, two million mock data sets have been generated from the on- and off-plane \xmax{} distributions of the in-FidFoV data set. The non-flat acceptance and lower \xmax{} resolution of the out-FidFoV data set has then been forward folded onto these samples to create out-FidFoV mock data sets which assume the difference in-FidFoV is real. The AD-test is then computed for each mock data set. The blue histogram represents the corresponding $TS$ distribution. It peaks exactly at the value observed with the out-FidFoV data set, showing that the out-FidFoV is indeed less sensitive to the tested anisotropy. Overall, the independent test with the out-FidFoV shows that the on/off separation is present in both data sets, and that the difference seen in the out-FidFoV sample is consistent with the magnitude of the difference seen in the in-FidFoV sample. 
\section{Mapping the UHECR sky in composition}

\begin{figure*}[!htb]
    \centering
    \includegraphics[width=.75\textwidth]{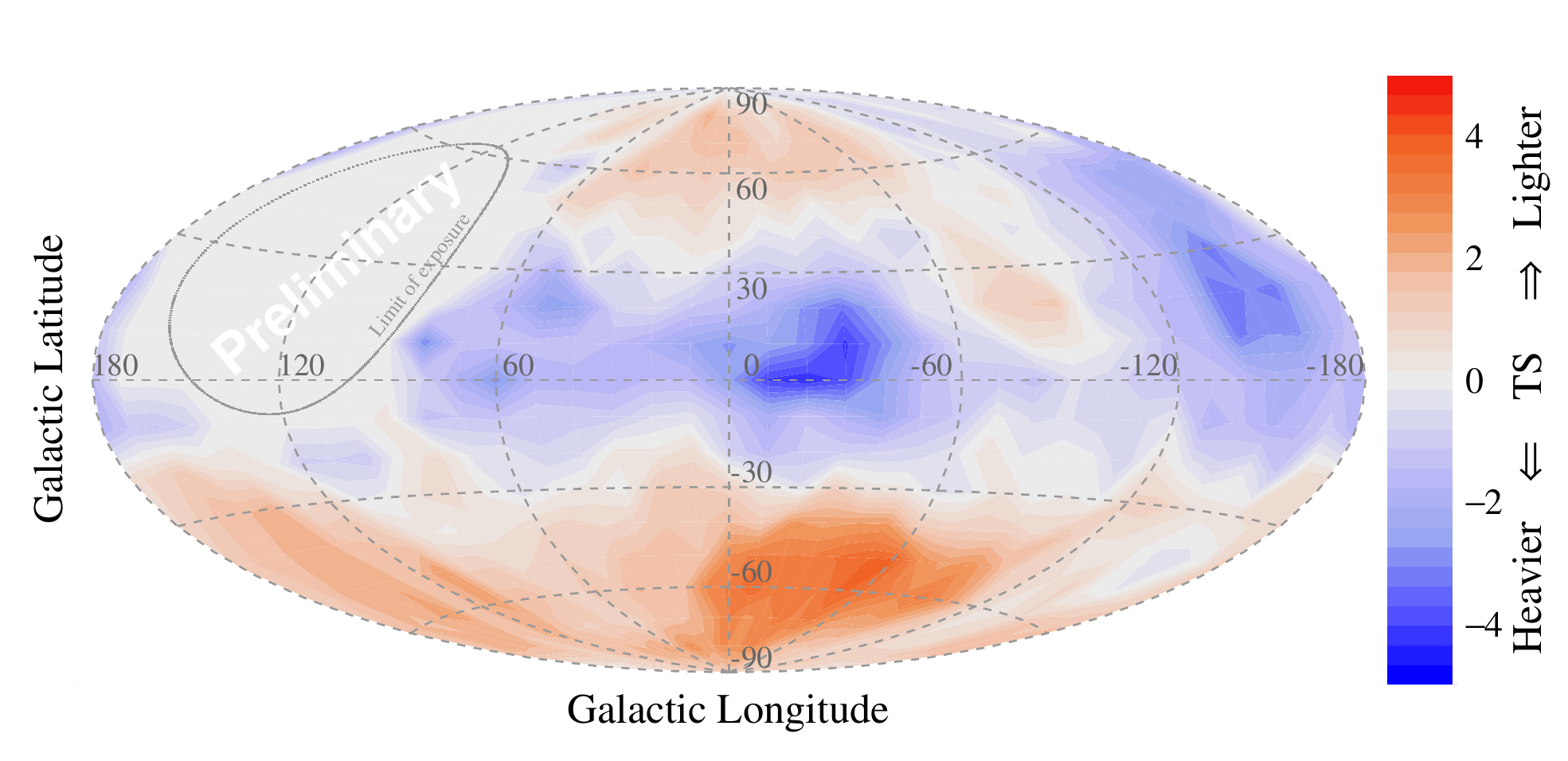}\vspace{-2mm}
    \caption{Sky map of comic ray composition for $E \geq 10^{18.7}$ eV}
    \label{fig:Map}
    \vspace{-1mm}
\end{figure*}

\vspace{-.1cm}
To aid interpretation of the latitude-dependent difference in composition, a test statistic quantifying the relative difference in \xmax{} between different parts of the sky is mapped in \autoref{fig:Map}, for UHECR primaries with $E \geq 10^{18.7}$\,eV. To produce this map from the in-FidFoV data set, first the requirement $E \geq 10^{18.7}$\,eV is imposed. Then,  because small portions of the sky are analysed, in contrast to the on/off study, each event has its $B$, $R$, and $A$ corrected for based on its arriving declination instead of its arriving galactic latitude. This is because local geometry has a time-independent relationship with arrival declination. This means $B$, $R$, and $A$ can be corrected equally well for each direction in the sky\footnote[1]{Using declination-dependent corrections changes the on/off comparison only by $+0.1$\,\gcm{} and increases systematic uncertainties.}. 

At this point, a top-hat sampling is used to collect all events with arrival directions within $30^\circ$ of a point $(\ell,b)$ into an \textit{in-hat} sample. All other events are placed in an \textit{out-hat} sample. The distributions of \xmaxnorm{} for the in-hat and out-hat samples are then compared using Welch's t-test~\cite{welch1938significance}:
\begin{equation}
    TS = \frac{\langle X_{\text{max}}^{\prime\,\text{in}} \rangle - \langle X_{\text{max}}^{\prime\,\text{out}} \rangle}{\sqrt{\left(\sigma\left( X_{\text{max}}^{\prime\,\text{in}} \right)/\sqrt{N^{in}}\right)^2 + \left(\sigma\left( X_{\text{max}}^{\prime\,\text{out}} \right)/\sqrt{N^{out}}\right)^2}},
\end{equation}
where $N^{in}$ and $N^{out}$ are the event counts for the in- and out-hat samples respectively\footnote{Because Welch's t-test considers event statistics, the FD arrival direction-dependent exposure is naturally treated through its use.}. This procedure is repeated for top-hats centered on each point in a $5^\circ$ by $5^\circ$ galactic latitude and longitude grid. The result is shown in \autoref{fig:Map}, which illustrates the relative composition of UHECRs with $E\geq 10^{18.7}$\,eV arriving from each point in the sky.

In \autoref{fig:Map}, positive $TS$ values (red) indicate that events within $30^\circ$ of that point have a lighter mean mass than the rest of the sky. Negative values (blue) indicate that events within $30^\circ$ of that point have a heavier mean mass than the rest of the sky. An excess of heavy particles within $30^\circ$ of the galactic plane is visible. This can not be  due to detector systematics as they would be declination dependent and appear as radial patterns centered on $\ell = -57^{\circ}, b =-27^{\circ}$ due to the geographic location of the Observatory. 
\section{Conclusions and Outlook}
There is an apparent difference in the mean mass of primaries with energies greater than $10^{18.7}$\,eV that arrive from within $30^\circ$ of the galactic plane. This has been observed at least at the $3.3\,\sigma$ level in the standard hybrid data set used for $X_{\text{max}}$-based composition analyses. It has now been independently confirmed with an additional $2.1\,\sigma$ significance in a second hybrid data set formed from high quality events cut by a selection aimed at reducing the bias caused by the $X_{\text{max}}$-dependent event acceptance. The combined significance of these two results has not yet been evaluated. Further tests of the on-/off-plane difference are being planned using analyses of data from the SD and will be reported elsewhere. 

Currently, this result should be considered to primarily provide a new verification of a mixed composition above the ankle as it is clear no such difference could be observed in a flux with a  single mass component. Though the analysis provides a possible indication that the galactic magnetic field may have an observable impact on mass-dependent anisotropies, the result found in this analysis does not necessarily support a causal relationship with galactic structures. The differing horizons of different nuclear species at a given energy could also result in composition-dependent anisotropic patterns~\cite{Ding:2021emg}. It is important, however, to note there is significant tension with models~\cite{Allard:2021ioh}. Alternative scenarios are being explored along these lines of thought.

\bibliography{HeadersAndConfig/References}{}

\end{document}